\documentclass[aps,pra,reprint,amsmath,amssymb,groupedaddress,showpacs]{revtex4-1}
\pdfoutput=1                          
\usepackage{graphicx}                  
\usepackage{float}
\usepackage{dcolumn}                   
\usepackage{bm}                        
\usepackage[bookmarks=false]{hyperref} 
\usepackage[font={small}]{caption}     
\usepackage{enumitem}                  
\usepackage{textcomp}                  
\hypersetup{
    unicode=false,                     
    pdftoolbar=true,                   
    pdfmenubar=true,                   
    pdffitwindow=true,                 
    pdfstartview={},                   
    pdftitle={Multipartite Mixed Maximally Entangled States: Mixed States with Entanglement 1},                       
    pdfauthor={Samuel R. Hedemann},    
    pdfsubject={Multipartite Entanglement},
    pdfcreator={},                     
    pdfproducer={},                    
    pdfkeywords={Mixed Maximal Entanglement,} {MME,} {Entanglement,} {Multipartite,} {True-Generalized X States,} {TGX States,} {TGX,} {Multipartite Schmidt Decomposition,} {Ent},         
    pdfnewwindow=true,                 
    colorlinks=true,                   
    linkcolor=black,                   
    citecolor=black,                   
    filecolor=black,                   
    urlcolor=black                     
}
\newcommand{\Sec}[1]{\hyperref[sec:#1]{Sec.{\kern 2pt}\ref*{sec:#1}}}
\newcommand{\Section}[1]{\hyperref[sec:#1]{Section~\ref*{sec:#1}}}
\newcommand{\Secs}[2]{\protect\hyperref[sec:#1]{{Secs.{\kern 2pt}\ref*{sec:#1}--\ref*{sec:#2}}}}
\newcommand{\Fig}[2][]{\hyperref[fig:#2]{Fig.{\kern 2pt}\ref*{fig:#2}#1}}
\newcommand{\Figure}[2][]{\hyperref[fig:#2]{Figure~\ref*{fig:#2}#1}}
\newcommand{\App}[1]{\hyperref[sec:App.#1]{App.{\kern 2pt}\ref*{sec:App.#1}}}
\newcommand{\Appendix}[1]{\hyperref[sec:App.#1]{Appendix~\ref*{sec:App.#1}}}
\newcommand{\EqLab}[1]{\\\noindent\smash{\raisebox{6pt}[0pt][0pt]{\hypertarget{eq:#1}{}}}\vspace{-11pt}}
\newcommand{\Eq}[1]{\protect\hyperlink{eq:#1}{(\ref*{eq.#1})}}
\newcommand{\Eqs}[2]{\protect\hyperlink{eq:#1}{(\ref*{eq.#1}--\ref*{eq.#2})}}
\newenvironment{Equation}[1]{\EqLab{#1}\begin{equation}\label{eq.#1}}{\end{equation}\par\noindent\ignorespacesafterend}
\newcommand{\Table}[2][]{\hyperref[tab:#2]{Table~\ref*{tab:#2}#1}}
\newcommand{\Tables}[2]{\hyperref[tab:#1]{Tables~\ref*{tab:#1}--\ref*{tab:#2}}}
\newcommand{\ThmLab}[1]{\noindent\smash{\raisebox{11pt}[0pt][0pt]{\hypertarget{Thm:#1}{}}}\textbf{Theorem{\kern 3pt}#1:~~}}
\newcommand{\Thm}[1]{\protect\hyperlink{Thm:#1}{Theorem{\kern 3pt}#1}}
\newcommand{\Thms}[2]{\protect\hyperlink{Thm:#1}{Theorems{\kern 3pt}#1--#2}}
\newcommand{\tr}[0]{\text{tr}}
\newcommand{\shiftmath}[2]{\textnormal{\raisebox{#1}[#1][#1]{$#2$}}}
\newcommand{\scalemath}[2]{\textnormal{\scalebox{#1}{$#2$}}}
\newcommand{\hsp}[1]{{\kern #1pt}}
\renewcommand{\geq}[0]{\geqslant}
\renewcommand{\ge}[0]{\geqslant}
\renewcommand{\leq}[0]{\leqslant}

\newcommand{\OneOverSqrt}[1]{{\scalemath{0.80}{\frac{1}{\rule{0pt}{8pt}\sqrt{\rule{0pt}{5pt}#1}}}}}
\newcommand{\redx}[2]{{#1}{\kern -5.3pt}{{\textnormal{\raisebox{-1.2pt}{\scalebox{1.2}{\textasciicaron}}}}}{\kern -4.2pt}{~}^{(#2)}}
\newcommand{\redxsuper}[3]{{#1}{\kern -5.3pt}{{\textnormal{\raisebox{-1.2pt}{\scalebox{1.2}{\textasciicaron}}}}}{\kern -4.2pt}{~}^{(#2)#3}}
\newcommand{\mbar}[0]{{{\kern 0.7pt}\overline{{\kern -0.7pt}m\rule{0pt}{4.8pt}{\kern -0.9pt}}{\kern 0.9pt}}}
\newcommand{\mbarsub}[0]{{\kern 0.0pt}\mathop {m}\limits^{{\kern -0.3pt}{\overline{{\kern 5.5pt}}}}{\kern 0.0pt}}
\newcommand{\nmaxnot}{n_{\,\overline{{\kern -1.8pt}\max^{~^{~^{~}}}\!\!\!\!\!\!\!\!\!\!}}\,}

\newcommand{\TGX}[0]{{\text{TGX}}}
\newcommand{\ME}[0]{{\text{ME}}}
\newcommand{\MME}[0]{{\text{MME}}}

\newcommand{\LU}[0]{{\text{LU}}}
\newcommand{\EPU}[0]{{\text{EPU}}}
\newcommand{\specialstar}[0]{{\scalemath{1.1}{\star}}}
\newcommand{\maxMMErank}[0]{{R_{\MME}}}
\newcommand{\maxMMEranklim}[0]{{R_{\MME}^{\lim}}}
\begin{document}
\title{Multipartite Mixed Maximally Entangled States: Mixed States with Entanglement 1}
\author{Samuel R. Hedemann}
\noaffiliation
\date{September 23, 2021}
\begin{abstract}
We present a full definition of mixed maximally entangled (MME) states for multipartite systems, generalizing their existing definition for bipartite systems by using multipartite Schmidt decomposition. MME states are a special kind of maximally entangled mixed state (MEMS) for which every pure decomposition state in all decompositions is maximally entangled. Thus, MME states have entanglement 1 by all valid unit-normalized entanglement measures, whereas general MEMS can have entanglement less than 1. Multipartite MME states likely have important applications such as remote state preparation, and also set critical performance goals for entanglement measures.
\end{abstract}
\maketitle
\section{\label{sec:I}Introduction: Bipartite MME}
In \cite{LZFF}, the amazing phenomenon that there exist mixed states with the same entanglement \cite{Schr,EPR} as maximally entangled (ME) pure states was presented and fully characterized for bipartite systems. Such \textit{mixed maximally entangled} (MME) states have the property that \textit{all} pure decomposition states of \textit{all} of their decompositions are maximally entangled by any valid measure of entanglement, so those pure states' reductions are both as simultaneously mixed as it is possible for them to be, given a pure parent state.  (We also use MME to mean \textit{mixed maximal entanglement}; the property of a mixed state having the entanglement of an ME pure state.)

\textit{Mixed states} have form \smash{$\rho\!=\!\sum\nolimits_j p_{j}\rho_{j}$} where $\rho_{j}\!\equiv\! |\psi_{j}\rangle\langle\psi_{j}|$ is a \textit{pure decomposition state}, $p_j\!\in\![0,1]\;\forall j$, and \smash{$\sum\nolimits_{j}p_{j}\!=\!1$}. We often treat \textit{pure states} $\rho=|\psi\rangle\langle\psi|$ as special cases of mixed states where $p_k =1$ and $p_{j\neq k}=0$, allowing $\rho$ to represent \textit{all} states. Similarly, although we will mainly speak of MME as a novel effect of \textit{strictly mixed states} (nonpure states), we will sometimes find it useful to use pure-state-inclusive characterizations of MME.

An example of MME states in a $2\times 4$ system is
\begin{Equation}                      {1}
\rho _{\MME}  =\! \frac{1}{2}\!\left( {\begin{array}{*{20}c}
   {\lambda _1} &  \cdot  &  \cdot  &  \cdot  &  \cdot  &  \cdot  &  \cdot  & {\lambda _1}  \\
    \cdot  & {\lambda _2} &  \cdot  &  \cdot  &  \cdot  &  \cdot  & {\lambda _2} &  \cdot   \\
    \cdot  &  \cdot  &  \cdot  &  \cdot  &  \cdot  &  \cdot  &  \cdot  &  \cdot   \\
    \cdot  &  \cdot  &  \cdot  &  \cdot  &  \cdot  &  \cdot  &  \cdot  &  \cdot   \\
    \cdot  &  \cdot  &  \cdot  &  \cdot  &  \cdot  &  \cdot  &  \cdot  &  \cdot   \\
    \cdot  &  \cdot  &  \cdot  &  \cdot  &  \cdot  &  \cdot  &  \cdot  &  \cdot   \\
    \cdot  & {\lambda _2} &  \cdot  &  \cdot  &  \cdot  &  \cdot  & {\lambda _2} &  \cdot   \\
   {\lambda _1} &  \cdot  &  \cdot  &  \cdot  &  \cdot  &  \cdot  &  \cdot  & {\lambda _1}  \\
\end{array}} \right)\!,
\end{Equation}
or {$\rho _{\MME}  \hsp{-1}=\hsp{-1} \lambda _1 |\Phi _{1,8}^{+}  \rangle \langle \Phi _{1,8}^{+}  | \!+\! \lambda _2 |\Phi _{2,7}^{+}  \rangle \langle \Phi _{2,7}^{+}  |$} where $\lambda_k \!\in\! (0,1)$ are \textit{any} normalized eigenvalues, \smash{$|\Phi _{1,8}^{+}\rangle\equiv \shiftmath{2pt}{\OneOverSqrt{2}}(|1\rangle+|8\rangle)=$} \smash{$\shiftmath{2pt}{\OneOverSqrt{2}}(|1,1\rangle+|2,4\rangle)$} and \smash{$|\Phi _{2,7}^{+}\rangle\equiv \shiftmath{2pt}{\OneOverSqrt{2}}(|2\rangle+|7\rangle)=\shiftmath{2pt}{\OneOverSqrt{2}}(|1,2\rangle+$} \smash{$|2,3\rangle)$} are ME pure states where $|a\rangle$ is the $a$th computational basis state where labels start on $1$ so $a\in 1,\ldots,8$, and in the coincidence basis $|a\rangle=|a_{1},a_{2}\rangle\equiv|a_{1}\rangle|a_{2}\rangle\equiv|a_{1}\rangle\!\otimes\!|a_{2}\rangle$, with $a_{1}\!\in\! 1,2$ and $a_{2}\!\in\! 1,\ldots,4$. Most notably,
\begin{Equation}                      {2}
E(\rho _{\MME} )=1
\end{Equation}
for \textit{all} normalized valid measures of entanglement $E(\rho)$.

However, \textit{MME states are not possible in all systems}. As explained in \cite{LZFF}, for bipartite mixed states $\rho\equiv\rho^{(1,2)}$ in a Hilbert space \smash{$\mathcal{H}\equiv\mathcal{H}^{(1)}\otimes\mathcal{H}^{(2)}$} of dimension $n\!\equiv\!\dim(\rho)\!=\! n_{1} n_{2}$, where \smash{$n_{m}$} is the dimension of mode $m$ of $\mathcal{H}^{(m)}$, MME states of rank $r\equiv\text{rank}(\rho)$ in such an $n_1  \times n_2$ system are possible whenever $n_B \geq rn_S|_{r\geq 2}$, so
\begin{Equation}                      {3}
2\leq r\leq {\textstyle \text{floor}\left({\frac{n_{B}}{n_{S}}}\right)},
\end{Equation}
where \smash{$n_S \!\equiv\!\min\{\mathbf{n}\}$}, \smash{$n_B \!\equiv\! \max\{\mathbf{n}\}\!=\!n/n_S$}, \smash{$\mathbf{n}\equiv (n_1 ,n_2 )$}, \smash{$n = n_1 n_2$}, and $r\geq 2$ is the minimal requirement for $\rho$ to be strictly mixed. Thus, \textit{the condition for a bipartite system to be capable of hosting MME}, a property we call \textit{bipartite MME potential}, is $n_B \geq 2n_S$, so
\begin{Equation}                      {4}
{\textstyle \text{floor}\left({\frac{n_{B}}{n_{S}}}\right)}\geq 2,
\end{Equation}
since otherwise, the upper limit in \Eq{3} would be below the lower limit, and no $r\geq 2$ could satisfy it. (When $r=1$, $n_B \geq rn_S$ still holds, but then any maximally entangled states are \textit{pure}, so they are simply  ME pure states, which are possible in all multipartite systems.)

Here, we extend the idea of MME states to $N$-partite systems, for mixed states represented by density operators $\rho\equiv\rho^{(\mathbf{N})}\equiv\rho^{(1,\ldots,N)}$, where parenthetical superscripts are labels of subsystems (modes) of an $N$-partite Hilbert space \smash{$\mathcal{H}\equiv\mathcal{H}^{(\mathbf{N})}\equiv\mathcal{H}^{(1)}\otimes\cdots\otimes\mathcal{H}^{(N)}$} of dimension $n\!\equiv\!\dim(\rho)\!=\! n_{1}\cdots n_{N}$, where \smash{$n_{m}$} is the dimension of mode $m$ of Hilbert space $\mathcal{H}^{(m)}$, and $\mathbf{n}\equiv(n_{1},\ldots,n_{N})$. 

The type of multipartite entanglement we will use is \textit{full $N$-partite entanglement}, meaning states that do \textit{not} have an optimal decomposition of the $N$-partite separable form \smash{$\rho=\sum\nolimits_j p_{j}\rho_{j}^{(1)}\otimes\cdots\otimes\rho_{j}^{(N)}$} where \smash{$\rho_{j}^{(m)}$} are pure states of mode $m$ and $p_j\in[0,1]\;\forall j$ where \smash{$\sum\nolimits_{j}p_{j}=1$} and $N\geq 2$. These results can then be generalized to other forms of \textit{distinctly multipartite entanglement} via the methods in \cite{HedC,HCor}, since full $N$-partite entanglement is a building-block of such generalizations.

To make these new results accessible, \Sec{II} is a brief \textit{summary of results for multipartite MME}, and the sections that follow give derivations and proofs of those results.  \Section{III} reviews the \textit{multipartite Schmidt decomposition} first presented in \cite{HedE}, which enables \Sec{IV} to present the \textit{multipartite MME conditions} generalizing the idea of MME to multipartite systems.  Then, \Sec{V} shows \textit{tests and examples of multipartite MME} for a variety of systems. In the Conclusions in \Sec{VI} we mention the impacts of MME on entanglement measures and suggest applications such as remote state preparation.
\section{\label{sec:II}Summary of Results for Multipartite MME}
The ability of any $N$-partite system of mode structure $\mathbf{n}\equiv(n_{1},\ldots,n_{N})$ to host MME states is its \textit{MME potential}, indicated by its \textit{maximal MME rank} $\maxMMErank$ as
\begin{Equation}                      {5}
\setlength\fboxsep{4pt}   
\setlength\fboxrule{0.5pt}
\fbox{$
\maxMMErank\equiv\!\mathop {\max \{ r\}; }\limits_{r\,\in\, 1,\ldots,\maxMMEranklim}\text{s.t.}\left\{\!\! {\begin{array}{*{20}l}
   {{ \maxMMEranklim \equiv \text{floor}}\!\left(\! {\textstyle \frac{{\min \{ \mathbf{n}_B \} }}{{\min \{ \mathbf{L}_* \} }}}\!\right)\!;}  \\
   {\exists \{ |\Phi _k \rangle \}_{k = 1}^{r}\!\in\!\{|\Phi _{\ME_{\TGX} } \rangle\};\rule{0pt}{10pt}}  \\
   {\tr[\tr_{B_m }\! (|\Phi _k \rangle{\kern -1pt} \langle \Phi _l |)]\! =\! \delta _{k,l}\,\,{\forall m}\rule{0pt}{12pt}}  \\
\end{array}} \right.\!\!\!$}
\end{Equation}
where \smash{$|\Phi _{\ME_{\TGX} } \rangle$} is a special kind of ME state reviewed in \Sec{III.A}, \smash{$\mathbf{n}_B  \!\equiv\! (n_{B_1 } , \ldots ,n_{B_N } )$}, \smash{$n_{B_m }  \!\equiv\! \max \{ n_m ,n_{\mbarsub} \}$}, \smash{$n_{\mbarsub}\!\equiv\!\frac{n}{n_{m}}$}, $B_m$ is the label for mode(s) of dimension $n_{B_m } $ for extreme bipartition $(m|\mbar)$, for modes $m\in 1,\ldots,N$, where $\mbar \equiv \{ 1, \ldots ,m - 1,m + 1, \ldots ,N\} $, and $\mathbf{L}_*  \equiv \{ L_* \}  \equiv \{ L_{*1} ,L_{*2} , \ldots \} $ are the values of $L$ that satisfy
\begin{Equation}                      {6}
\mathop {\min }\limits_{L \in 2, \ldots ,\nmaxnot } \left[{\scalemath{0.85}{\frac{1}{N}\sum\limits_{m = 1}^N {\frac{{n_m P_{\text{MP}}^{(m)} (L) - 1}}{{n_m  - 1}}}}}\right]\!,
\end{Equation}\\
where \smash{$\nmaxnot\equiv \frac{n}{n_{\max}}$}, \smash{$n_{\max}\equiv\max\{\mathbf{n}\}$}, and\\
\vspace{-12pt}
\begin{Equation}                      {7}
\begin{array}{*{20}l}
   {P_{\text{MP}}^{(m)} (L)\equiv} &\!\! { \bmod (L,n_m )\left( {\frac{{1 + \text{floor} (L/n_m )}}{L}} \right)^2 }  \\
   {} &\!\! { + [n_m  - \bmod (L,n_m )]\left( {\frac{{\text{floor}(L/n_m )}}{L}} \right)^2 \!,}  \\
\end{array}
\end{Equation}
\vspace{-6pt}\\
where $\bmod (a,b) \equiv a - [\text{floor}(a/b)]b$. Physical meanings of \Eqs{6}{7} are given in \cite{HedE} and discussed in \Sec{III}.

If $\maxMMErank\geq 2$, then the system can host MME for ranks 
\begin{Equation}                      {8}
2\leq r\leq R_{\MME},
\end{Equation}
whereas if $\maxMMErank=1$, the system can only host \textit{pure} ME states.\hsp{-2} Thus, \Eq{8} generalizes \Eq{3}, while \Eq{5} generalizes \Eq{4}.

\Figure{1} shows how an MME state in $2\times 2 \times 2 \times 2$ has entanglement $1$ for all decompositions by using the full $N$-partite entanglement measure \textit{the ent} $\Upsilon$ \cite{HedE}, and compares this to a nonMME entangled state and a separable state in the same system. The tables in \Sec{V} give examples of $\maxMMErank$ and show how to construct multipartite MME states for a variety of multipartite systems.

MME states are a special kind of maximally entangled mixed state (MEMS) \cite{IsHi,ZiBu,HoBM,VeAM,WNGK,MeMH}. MEMS are mixed states with the most entanglement possible for any state of their mixedness, but in general they have entanglement $E\leq 1$. 

Next, we present the derivation of the above results.
\begin{widetext}
{~}\vspace{-12pt}
\begin{figure}[H]
\centering
\includegraphics[width=1.00\textwidth]{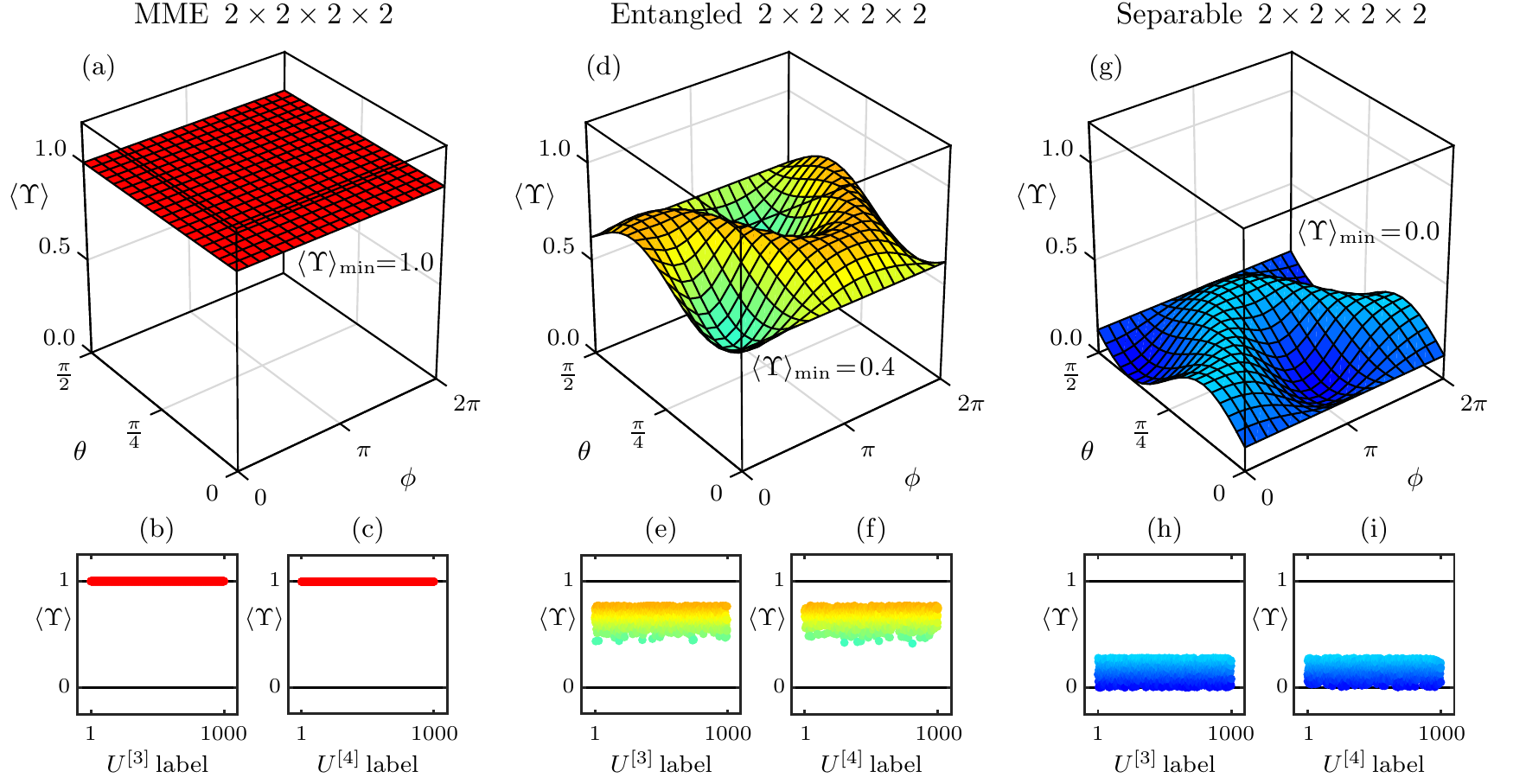}
\vspace{-13pt}
\caption{(color online) (a), (b), (c) show entanglement as minimum average ent \smash{$\langle\Upsilon\rangle_{\min}$} over many decompositions with decomposition unitaries \smash{$U^{[D]}\!\equiv\! U^{[D]}(\theta,\phi,\ldots)$} of dimensions $D\in\{r,\ldots,r^2\}=\{2,3,4\}$ for a rank-$2$ MME state \smash{$\rho_{\MME}$} in $2\times 2 \times 2 \times 2$ with arbitrary spectrum $\{\lambda_1,\lambda_2\}\!=\!\{0.7,0.3\}$ (\Secs{IV}{V} define $U$). Each grid point or dot is $\langle\Upsilon\rangle$ for a different decomposition of \smash{$\rho_{\MME}$}. This test strongly supports the claim that all pure decomposition states of all decompositions of \smash{$\rho_{\MME}$} have $\Upsilon\!=\!1$, so the minimum $\langle\Upsilon\rangle$ over all decompositions is \smash{$\langle\Upsilon\rangle_{\min}\!=\!1$}, and thus \smash{$\rho_{\MME}$} has the same entanglement as an ME pure state. For comparison, (d), (e), (f) show the same tests for an arbitrary rank-$2$ \textit{entangled nonMME} state (\smash{$0<\langle\Upsilon\rangle_{\min}\!<\!1$}) in $2\times 2 \times 2 \times 2$, and (g), (h), (i) test an arbitrary rank-$2$ \textit{separable} state (\smash{$\langle\Upsilon\rangle_{\min}\!=\!0$}) in $2\times 2 \times 2 \times 2$.  Identical MME plots were confirmed for each MME-hosting system in \Tables{1}{5}. These tests are not proofs; they are merely necessary tests of the proof in \Sec{IV}.}
\label{fig:1}
\end{figure}
\end{widetext}
\section{\label{sec:III}Multipartite Schmidt Decomposition}
The multipartite Schmidt decomposition (first presented in \cite{HedE}) requires that we first discuss a special family of states called \textit{true-generalized X (TGX) states}.  Therefore, here we first review TGX states, and then discuss the multipartite Schmidt decomposition in preparation for using it to define multipartite MME states.
\subsection{\label{sec:III.A}A Review of TGX States}
X states \cite{YuEb,WBPS,YE04,AlJa,Wein,PABJ} are an important type of state for two qubits. In \cite{HedX} we presented strong evidence that X states are universal for entanglement, which was later proved implicitly in \cite{MeMG} and proved explicitly in \cite{HeXU}, which showed that any general two-qubit mixed state can be unitarily transformed to an X state of the same spectrum and entanglement, a property called \textit{entanglement-preserving unitary (EPU) equivalence} (so-named because such  operations are generally nonlocal). Very important two-qubit X states are the Bell states, which are ME X states and form a complete orthonormal basis. In fact, their ability to form a basis is part of what allows ME X states to appear in the two-qubit Schmidt decomposition, a property we will discuss more below.

However, X states are \textit{not} generally the proper generalization of the compact EPU-equivalent states for full $N$-partite entanglement; the correct generalized states, proposed in \cite{HedX} and studied in \cite{HedD,MeMH,HedE,HedC,HeXU,HCor} appear to be the \textit{true-generalized X (TGX) states}, which are generally \textit{non}X-shaped, and are named for their ability to generalize the EPU equivalence of X states of two qubits to larger systems.  ME TGX states are the true multipartite generalization of the Bell states for full $N$-partite entanglement, and as shown in \cite{HedE}, they also enable a complete orthonormal basis which we call a \textit{maximally entangled basis (MEB)}. For \textit{A Brief History of TGX States}, see App.{\kern 2.5pt}H of \cite{HCor}. 

For example, general TGX states in $2\times 3$ are
\begin{Equation}                      {9}
\rho_{\TGX}\,  =\! \left( {\begin{array}{*{20}c}
   {\rho _{1,1} } &  \cdot  &  \cdot  &  \cdot  & {\rho _{1,5} } & {\rho _{1,6} }  \\
    \cdot  & {\rho _{2,2} } &  \cdot  & {\rho _{2,4} } &  \cdot  & {\rho _{2,6} }  \\
    \cdot  &  \cdot  & {\rho _{3,3} } & {\rho _{3,4} } & {\rho _{3,5} } &  \cdot   \\
    \cdot  & {\rho _{4,2} } & {\rho _{4,3} } & {\rho _{4,4} } &  \cdot  &  \cdot   \\
   {\rho _{5,1} } &  \cdot  & {\rho _{5,3} } &  \cdot  & {\rho _{5,5} } &  \cdot   \\
   {\rho _{6,1} } & {\rho _{6,2} } &  \cdot  &  \cdot  &  \cdot  & {\rho _{6,6} }  \\
\end{array}} \right)\!,
\end{Equation}
while ME TGX states in $2\times 3$ are
\begin{Equation}                      {10}
\begin{array}{*{20}l}
   {|\Phi _{1,5}^ \pm  \rangle } &\!\! { \equiv \frac{1}{{\sqrt 2 }}(|1\rangle  \pm |5\rangle ),} &\;\, {|\Phi _{1,6}^ \pm  \rangle } &\!\! { \equiv \frac{1}{{\sqrt 2 }}(|1\rangle  \pm |6\rangle ),}  \\
   {|\Phi _{2,6}^ \pm  \rangle } &\!\! { \equiv \frac{1}{{\sqrt 2 }}(|2\rangle  \pm |6\rangle ),} &\;\, {|\Phi _{2,4}^ \pm  \rangle } &\!\! { \equiv \frac{1}{{\sqrt 2 }}(|2\rangle  \pm |4\rangle ),}  \\
   {|\Phi _{3,4}^ \pm  \rangle } &\!\! { \equiv \frac{1}{{\sqrt 2 }}(|3\rangle  \pm |4\rangle ),} &\;\, {|\Phi _{3,5}^ \pm  \rangle } &\!\! { \equiv \frac{1}{{\sqrt 2 }}(|3\rangle  \pm |5\rangle ),}  \\
\end{array}
\end{Equation}
\cite{HedX}, which have the property that both reductions are as simultaneously mixed as they can be for a pure parent state, as explained in \cite{HedX,HedE}. Note that two different MEBs can be formed from \Eq{10} (grouped by columns).

We call the set of scalar indices in a TGX state an \textit{ME TGX tuple}, so the first ME TGX tuple in \Eq{10} is $\{1,5\}$.

The \textit{number of indices} in an ME TGX tuple is $L_*$, and is the number of levels with nonzero probability amplitudes of equal magnitude supporting full $N$-partite ME.  Generally, multiple different $L_*$ values can exist in a given system, so we represent them as $\mathbf{L}_*\equiv\{L_*\}$ (see Table II of \cite{HedE} for $L_*$ values of various multipartite systems).

The $L_*$ values of any multipartite system are calculated from its \textit{minimum physical simultaneous reduction purities (MPSRPs)}, meaning the collection of purities of the single-mode reductions of an ME state. The single-mode reduction of $\rho$ for mode $m$ is $\redx{\rho}{m}\equiv\tr_{\mbarsub}(\rho)$ which is the partial trace over all modes \textit{except} $m$, and by \textit{single modes} we mean the $N$ subsystems in which no coincidence events occur, that define the $N$-mode system. Thus, $\redx{\rho}{m}$ has dimension $n_m$.  \textit{Purity} is $P(\rho)\equiv\tr(\rho^{2})$, and the \textit{mode-$m$ reduction purity} is $P^{(m)}\equiv P^{(m)}(\rho)\equiv\tr[(\redx{\rho}{m})^{2}]$. For a \textit{lone} $n_m$-level system, the minimum purity is $\frac{1}{n_m}$, but in a multipartite system, the reduction purities cannot always reach that ideal minimum, depending on mode sizes and parent state $\rho$, requiring that we specify \textit{physical} reduction purities, as in \cite{HedE}.

We gauge how close a state is to achieving MPSRPs by first identifying the slightly less-general quantities \smash{$P_{\text{MP}}^{(m)}(L)$} \textit{as functions of $L$ levels of equal nonzero probability amplitudes of a pure parent TGX state}, given in \Eq{7}.  Then, to find a measure that indicates the lowest simultaneous purity all mode-$m$ reductions can have given a pure parent state, we remap each general $P^{(m)}$ to $[0,1]$, take the arithmetic mean over $N$ modes, and find all values of $L\in 2,\ldots,\nmaxnot$ that minimize that mean, designating them as $L_*$ values, as shown in \Eq{6}, so \smash{$P_{\text{MP}}^{(m)}(L_*)$} is the mode-$m$ MPSRP. This is the process for explicitly normalizing the \textit{ent}, as derived in \cite{HedE}.

ME TGX states, as in \Eq{10}, can be methodically generated by the 13-step algorithm from \cite{HedE}, which, for a given $L_*$ value and starting level, outputs a matrix $L_{\ME}$ whose rows are ME TGX tuples. In $2\times 3$, all such outputs yield
\begin{Equation}                      {11}
L_{\ME}\,=\left( {\begin{array}{*{20}c}
   1 & 5  \\
   1 & 6  \\
   2 & 4  \\
   2 & 6  \\
   3 & 4  \\
   3 & 5  \\
\end{array}} \right)\!.
\end{Equation}

As we will see, it is very useful to convert readily between the scalar-index representation of outcomes (levels) and the vector-index form of the coincidence basis.  For this, we use the \textit{inverse indical register function} [which maps scalar indices to vectors, such as $1\to\{1,1\}$ and $6\to\{2,3\}$ for converting \smash{$\frac{1}{{\sqrt 2 }}(|1\rangle  \pm |6\rangle )$} to \smash{$\frac{1}{{\sqrt 2 }}(|1,1\rangle  \pm |2,3\rangle )$}], and the \textit{indical register function}, such as $\{2,3\}\to 6$. Both of these functions were presented in \cite{HedE} and compactly summarized in App.{\kern 2.5pt}U of \cite{HCor}.

Here, we are most interested in the role of TGX states in multipartite Schmidt decomposition of pure states, which we now discuss.
\subsection{\label{sec:III.B}The Role of TGX States in Multipartite Schmidt Decomposition}
\vspace{-4pt}
From \cite{HedE}, a pure TGX state of arbitrary full $N$-partite entanglement in an $N$-partite system is
\vspace{-2pt}
\begin{Equation}                      {12}
\begin{array}{*{20}l}
   {|\Phi _j (\bm{\theta} )\rangle } &\!\! { \equiv \sum\limits_{h = 1}^{L_* } {x_h^{[L_* ]} (\bm{\theta} )|(L_{\ME} )_{j,h} \rangle } }  \\
   {} &\!\! { = \sum\limits_{h = 1}^{L_* } {x_h^{[L_* ]} (\bm{\theta} )|v_{j,h}^{(1)} \rangle  \otimes  \cdots  \otimes |v_{j,h}^{(N)} \rangle, } }  \\
\end{array}
\end{Equation}
\vspace{-8pt}\\
where $L_*$ is the number of levels in an ME TGX state (so it is the number of indices in an ME TGX tuple), \smash{$x_h^{[L_* ]} (\bm{\theta} )$} are Schl{\"a}fli's unit-hyperspherical coordinates of $L_*$ dimensions where $\bm{\theta}  \equiv (\theta _1 , \ldots ,\theta _{L_*  - 1} )\in [0,\frac{\pi }{2}]$ \cite{Schl}, \smash{$L_{\ME}  \equiv L_{\ME}^{\{ L_* \} } $} is the matrix output of the 13-step algorithm of \cite{HedE} whose row vectors are scalar-index ME TGX tuples, and \smash{$\mathbf{v}_{j,h} \equiv \{v_{j,h}^{(1)} , \ldots ,v_{j,h}^{(N)} \} \equiv \mathbf{v}_{\shiftmath{0pt}{(L_{\ME} )_{j,h}} }^{\shiftmath{1.3pt}{\{ N,\mathbf{n}\}} } $} is the vector index of coincidence labels corresponding to scalar index $(L_{\ME})_{j,h}$, where \smash{$\mathbf{a}_v^{\{ N,\mathbf{n}\} } $} is the inverse indical register function (App.{\kern 2.5pt}H of \cite{HedE}) which converts scalar index $v$ to vector index $\mathbf{a}$. Note that \smash{$v_{j,h}^{(m)}$} are integers and \smash{$|v_{j,h}^{(m)} \rangle$} is a mode-$m$ computational basis state, labeled as $1,\ldots,n_m$ by convention. Thus, $\mathbf{v}_{j,h}$ is the \textit{coincidence form} of the row-$j$, column-$h$ element of $L_{\ME}$.

For example, in $2\times 2$, $L_{\ME}=(\vphantom{|}_{2\hsp{5}3}^{1\hsp{5}4})$ since it has ME TGX tuples $\{1,4\}$ and $\{2,3\}$ for ``phaseless'' ME TGX states \smash{$|\Phi^{+}\rangle=\frac{1}{\sqrt{2}}(|1\rangle+|4\rangle)$} and \smash{$|\Psi^{+}\rangle=\frac{1}{\sqrt{2}}(|2\rangle+|3\rangle)$}, so
\begin{Equation}                      {13}
\begin{array}{*{20}l}
   {\left(\! {\begin{array}{*{20}c}
   1 & 4  \\
   2 & 3  \\
\end{array}}\! \right) \to \left(\! {\begin{array}{*{20}l}
   {\mathbf{v}_{1,1} } & {\mathbf{v}_{1,2} }  \\
   {\mathbf{v}_{2,1} } & {\mathbf{v}_{2,2} }  \\
\end{array}}\! \right)} &\!\! { = \left(\! {\begin{array}{*{20}c}
   {\{ v_{1,1}^{(1)} ,v_{1,1}^{(2)} \}} &\; {\{ v_{1,2}^{(1)} ,v_{1,2}^{(2)} \}}  \\
   {\{ v_{2,1}^{(1)} ,v_{2,1}^{(2)} \}} &\; {\{ v_{2,2}^{(1)} ,v_{2,2}^{(2)} \}}  \\
\end{array}}\! \right)}  \\
   {} &\!\! { = \left(\! {\begin{array}{*{20}c}
   {\{ 1,1\}} &\; {\{ 2,2\}}  \\
   {\{ 1,2\}} &\; {\{ 2,1\}}  \\
\end{array}}\! \right)\!,}  \\
\end{array}
\end{Equation}
which lets us specify the ME TGX coincidence expansion,
\begin{Equation}                      {14}
\begin{array}{*{20}l}
   {\frac{1}{{\sqrt 2 }}(|1\rangle  + |4\rangle )} &\!\! { = \frac{1}{{\sqrt 2 }}(|1,1\rangle  + |2,2\rangle )}  \\
   {\frac{1}{{\sqrt 2 }}(|2\rangle  + |3\rangle )} &\!\! { = \frac{1}{{\sqrt 2 }}(|1,2\rangle  + |2,1\rangle ).}  \\
\end{array}
\end{Equation}
Thus, in terms of the \textit{scalar} index \textit{argument} to \smash{$\mathbf{v}_{j,h}\equiv \mathbf{v}_{\shiftmath{0.5pt}{(L_{\ME} )_{j,h}} }^{\shiftmath{1.3pt}{\{ N,\mathbf{n}\}} }$} [which is $(L_{\ME})_{j,h}$], the $j$ references the $j$th ME TGX tuple [such as $\{2,3\}$ for $j=2$ in \Eq{13}], while $h$ references the $h$th element of that ME TGX tuple [such as $3$ for $h=2$ in ME TGX tuple $\{2,3\}$]. Then, $\mathbf{v}_{j,h}$ is the inverse indical register function applied to $(L_{\ME})_{j,h}$, for instance converting $3$ to $\{2,1\}$, as seen in \Eq{13}.

From \cite{HedE}, generally \textit{non}TGX states of the same entanglement as $|\Phi _j (\bm{\theta} )\rangle$ can be formed from $|\Phi _j (\bm{\theta} )\rangle$ by local unitary (LU) equivalence as
\begin{Equation}                      {15}
\begin{array}{*{20}l}
   {|\Phi _j '(\bm{\theta} )\rangle } &\!\! { \equiv (U^{(1)}  \otimes  \cdots  \otimes U^{(N)} )|\Phi _j (\bm{\theta} )\rangle }  \\
   {} &\!\! { = \sum\limits_{h = 1}^{L_* } {} x_h^{[L_* ]} (\bm{\theta} )|U_{:,v_{j,h}^{(1)} }^{(1)} \rangle  \otimes  \cdots  \otimes |U_{:,v_{j,h}^{(N)} }^{(N)} \rangle, }  \\
\end{array}
\end{Equation}
where $U^{(m)}$ is an $n_m$-level unitary operator in mode $m$, and $A_{:,c}$ is the $c$th column vector of matrix $A$ so that \smash{$|A_{:,c}\rangle$} is the ket formed from column $c$ of $A$; thus \smash{$|U_{\shiftmath{1.9pt}{\hsp{4}:,\hsp{2.5}v_{j,h}^{(m)}}}^{\shiftmath{0.8pt}{(m)}} \rangle\equiv U^{(m)}|v_{j,h}^{(m)} \rangle$}. (Note: as explained in \cite{HedE}, we could define an equivalent set of states with a generally \textit{nonlocal} diagonal unitary $D$ inserted in \Eq{15} as $(U^{(1)}  \otimes  \cdots  \otimes U^{(N)} )D|\Phi _j (\bm{\theta} )\rangle$, but the zeros of the TGX core $|\Phi _j (\bm{\theta} )\rangle$ cause $D$ to always have a form that allows it to be absorbed into the LU so that \Eq{15} is sufficient.)

The form in \Eq{15} is the \textit{multipartite Schmidt decomposition} (MSD) of a general pure state $|\psi\rangle\equiv|\Phi _j '(\bm{\theta} )\rangle$ in the context of full $N$-partite entanglement.  In general, systems posses multiple different $L_*$ values, so any given $|\psi\rangle$ may have multiple different MSDs, and since many different TGX cores are possible, multiple MSDs even exist when there is only one $L_*$ value.

The \smash{$x_h^{[L_* ]} (\bm{\theta} )$} are \textit{multipartite Schmidt coefficients}, and instead of requiring coincidence-index matching among the modes (such as in bipartite systems where the common Schmidt decomposition \cite{Schm} is \smash{$|\psi\rangle=\sum\nolimits_{q=1}^{n_{S}}{x_q^{[n_S ]}|U_{:,q }^{(1)} \rangle |U_{:,q }^{(2)} \rangle} $}\rule{0pt}{9.8pt} where $n_S \equiv\min\{\mathbf{n}\}$, since $L_{*}=n_{S}$ for $N=2$), here we use the more general \textit{TGX-indexing} which means that the indices must simply be the vector-indices of each level in any ME TGX tuple of the system. (We give a more general example in \Sec{V}, but in $2\times 2$ this shows that the Schmidt decomposition could also be written as $|\psi\rangle=$ \smash{$x_1^{[2 ]}|U_{:,1 }^{(1)} \rangle |U_{:,2 }^{(2)} \rangle +x_2^{[2 ]}|U_{:,2 }^{(1)} \rangle |U_{:,1 }^{(2)} \rangle$}\rule{0pt}{9.8pt} which is based on the ME TGX tuple $\{2,3\}=\{\{1,2\},\{2,1\}\}$\rule{0pt}{9.8pt}, rather than the more popular \smash{$|\psi\rangle=x_1^{[2 ]}|U_{:,1 }^{(1)} \rangle |U_{:,1 }^{(2)} \rangle +x_2^{[2 ]}|U_{:,2 }^{(1)} \rangle |U_{:,2 }^{(2)} \rangle$}\rule{0pt}{9.8pt},  based on ME TGX tuple $\{1,4\}=\{\{1,1\},\{2,2\}\}$\rule{0pt}{9.8pt}.)

From \cite{HedE}, any pure state is maximally full-$N$-partite entangled iff \smash{$x_{\hsp{1.5}h}^{\shiftmath{0pt}{[L_* ]}}  = \frac{1}{{\sqrt {L_* } }}\;\forall h\in 1\ldots,L_{*}$}, so that generally nonTGX ME pure states have the form
\vspace{-4pt}
\begin{Equation}                      {16}
\begin{array}{*{20}l}
   {|\Phi _j '\rangle} &\!\! {\equiv (U^{(1)}  \otimes  \cdots  \otimes U^{(N)} )|\Phi _j \rangle,}  \\
   {} &\!\! {= { \frac{1}{{\sqrt {L_* } }}}\sum\limits_{h = 1}^{L_* } {|U_{:,v_{j,h}^{(1)} }^{(1)} \rangle  \otimes  \cdots  \otimes |U_{:,v_{j,h}^{(N)} }^{(N)} \rangle },}  \\
\end{array}
\end{Equation}
so \textit{all ME states are LU equivalent to ME TGX states},
\begin{Equation}                      {17}
\begin{array}{*{20}l}
   {|\Phi _{\ME_{\TGX} } \rangle\equiv|\Phi _j \rangle } &\!\! { \equiv \frac{1}{{\sqrt {L_* } }}\sum\limits_{h = 1}^{L_* } {|(L_{\ME} )_{j,h} \rangle } }  \\
   {} &\!\! { = \frac{1}{{\sqrt {L_* } }}\sum\limits_{h = 1}^{L_* } {|v_{j,h}^{(1)} \rangle  \otimes  \cdots  \otimes |v_{j,h}^{(N)} \rangle , } }  \\
\end{array}
\end{Equation}
which reflects the fact proven in \cite{HedE} that all ME TGX states have $L_{*}$ nonzero probability amplitudes of equal magnitude. Since those probability amplitude magnitudes are exactly the multipartite Schmidt coefficients \smash{$x_h^{[L_* ]} (\bm{\theta} )$}, then the LU equivalence of all pure states to phaseless TGX states is why having $L_{*}$ multipartite Schmidt coefficients of equal values implies ME for \textit{non}TGX states as well, as stated before \Eq{16}. 

In the context of full $N$-partite entanglement, ME pure states are those for which all $N$ single-mode reductions have the lowest possible purity combination that they can simultaneously have given a pure parent state (minimum physical simultaneous reduction purities). (Note: as proven in \cite{HedE}, this does \textit{not} always mean that all reductions are maximally mixed states, \textit{nor} does it mean that their reductions are necessarily diagonal.)

Now we have the concepts we need to define multipartite MME, which we do next.
\section{\label{sec:IV}Multipartite MME Conditions}
Here we derive some necessary conditions for multipartite MME, and then identify some sufficient conditions and show that they are also necessary, yielding \Eq{5}.

By definition, \textit{all} decompositions of an MME state $\rho$ must contain \textit{only} ME pure states, which therefore \textit{includes} the \textit{spectral decomposition}. Therefore, a necessary condition for a rank-$r$ multipartite state $\rho$ to be MME is that it has spectral decomposition
\begin{Equation}                      {18}
\rho  = \sum\limits_{k = 1}^{r} {\lambda_k |\Phi _k '\rangle \langle \Phi _k '|},
\end{Equation}
where $\lambda_k  \in (0,1)$ are eigenvalues such that \smash{$\sum\nolimits_{k = 1}^{r} {\lambda_k  = 1}$}, and its eigenstates $|\Phi _k '\rangle$ are the generally \textit{non}TGX ME pure states of \Eq{16}. Again, by the definition of MME, as in \cite{LZFF}, any \textit{other} decomposition of $\rho$ must have the form
\begin{Equation}                      {19}
\rho  = \sum\limits_{j = 1}^{D \geq r} {p_j |w_j \rangle \langle w_j |} ,
\end{Equation}
where each $|w_j \rangle$ is \textit{also} ME, and given by
\begin{Equation}                      {20}
|w_j \rangle  \equiv \frac{1}{{\sqrt {p_j } }}\sum\limits_{k = 1}^{r} {U_{j,k} \sqrt {\lambda_k } |\Phi _k '\rangle } ,
\end{Equation}
where $U \equiv U^{[D \ge r]} $ is any $D$-level unitary, and
\begin{Equation}                      {21}
p_j  \equiv \sum\limits_{k = 1}^{r} {|U_{j,k} |^2 \lambda_k } ,\;\;\text{so also}\;\;\frac{1}{{p_j }}\sum\limits_{k = 1}^{r} {|U_{j,k} |^2 \lambda_k }  = 1.
\end{Equation}
Then, the single-mode reductions of each $|w_j \rangle$ are
\begin{Equation}                      {22}
\redx{\rho}{m}_{|w_j\rangle}  = \frac{1}{{p_j }}\sum\limits_{k = 1}^{r} {} \sum\limits_{l = 1}^{r} {} U_{j,k} U_{j,l} ^* \sqrt {\lambda_k \lambda_l } \tr_{\mbarsub} (|\Phi _k '\rangle \langle \Phi _l '|),
\end{Equation}
for $m\in 1,\ldots,N$ and $j \in 1, \ldots , D$ where $D \ge r$, and \smash{$\rho_{|\psi\rangle}\equiv|\psi\rangle\langle\psi|$}, so \smash{$\redx{\rho}{m}_{|\psi\rangle}\equiv\tr_{\mbarsub}(\rho_{|\psi\rangle})$}.

Since\rule{0pt}{11pt} MME requires \textit{all} $|w_j \rangle$ to be ME for any decomposition (a given $U$), an equivalent MME condition is
\begin{Equation}                      {23}
\redx{\rho}{m}_{|w_j\rangle}  = \redxsuper{\rho}{m}{\prime}_{\text{MP}_{j}}\;\;\;\forall m,\;\forall j,
\end{Equation}
where \smash{$\redxsuper{\rho}{m}{\prime}_{\text{MP}_{j}}$} is a mode-$m$ state of minimum physical simultaneous reduction purity (MPSRP) given a pure ME parent state [the prime is to match those in \Eq{22}].

Since MME requires \Eq{23} for \textit{all} decompositions (and hence all $U$), and since MPSRP states have specific reduction purities, another MME condition is that
\begin{Equation}                      {24}
P(\redx{\rho}{m}_{|w_j\rangle} )\;\;\text{is independent of }U\;\;\;\forall m,\;\forall j,
\end{Equation}
where again $P(\sigma)\equiv\tr(\sigma^{2})$ is \textit{purity}.  

Furthermore, to find the full set of MME states, we want to maximize the rank $r$ of $\rho$, so a side condition is
\begin{Equation}                      {25}
k\in 1,\ldots,\max\{r\}\;\;\text{s.t.}\;\;\rho\;\;\text{is MME}.
\end{Equation}

Further necessary conditions are implied by \Eq{18} as follows. First, note that since the generally nonTGX ME states $\{|\Phi _k '\rangle\}$ are orthonormal [since they are eigenstates of $\rho$ in \Eq{18}], they can be expressed as the first $r$ columns in a single unitary operator $W$ (where the remaining $n-r$ columns could be filled in by the maximally entangled basis (MEB) theorem from \cite{HedE}), so 
\begin{Equation}                      {26}
|\Phi _k '\rangle\equiv |W_{:,k} \rangle= W|k \rangle,
\end{Equation}
where $|k \rangle$ is the $k$th computational basis state.  Similarly, by the MEB, there always exists an orthonormal set of ME TGX states $\{|\Phi _k \rangle\}$ and some unitary $V$ such that
\begin{Equation}                      {27}
|\Phi _k \rangle\equiv |V_{:,k} \rangle= V|k \rangle,
\end{Equation}
so, inverting \Eq{27} as $|k \rangle=V^{\dag}|\Phi _k \rangle$, then \Eq{26} becomes
\begin{Equation}                      {28}
|\Phi _k '\rangle= WV^{\dag}|\Phi _k \rangle.
\end{Equation}
Thus, the MEBs $\{|\Phi _k '\rangle\}$ and $\{|\Phi _k \rangle\}$ are unitarily related. Furthermore, if $\{|\Phi _k '\rangle\}$ all have \textit{core ME TGX states} $\{|\Phi _k \rangle\}$ of the same $L_*$ in their MSD of \Eq{16}, then the unitary $WV^{\dag}$ in \Eq{28} is LU as
\begin{Equation}                      {29}
|\Phi _k '\rangle= U_{\LU}|\Phi _k 
\rangle=(U^{(1)}\otimes\cdots\otimes U^{(N)})|\Phi _k \rangle.
\end{Equation}
Thus, each $|\Phi _k '\rangle$ in \Eq{18} has a core ME TGX state $|\Phi _k \rangle$ preserving the orthonormality of $\{|\Phi _k '\rangle\}$, and if all $|\Phi _k \rangle$ have the same $L_*$, these sets are related by a \textit{single} LU, so there is no need to give index $k$ to $U_{\LU}$. (MME states with $\{|\Phi _k \rangle\}$ of \textit{multiple different} $L_*$ values may exist, but would be related to single-$L_*$ MME states by a nonlocal EPU, so we lose no generality by focusing on single-$L_*$ MME states in our quest to find all MME states.)

Now that we have some necessary conditions, we propose the following two mechanisms for achieving them as candidates for sufficient conditions for MME (they will require refining further below to achieve this sufficiency).%
\vspace{-1pt}
\begin{itemize}[leftmargin=*,labelindent=4pt]\setlength\itemsep{0pt}
\item[\textbf{1.}]\hypertarget{Mechanism:1}{}One way to achieve \Eq{24} \textit{and} \Eq{23} (at least for one $m$) would be if \smash{$U_{j,k} U_{j,l} ^* \sqrt {\lambda_k \lambda_l }$} gets simplified to $|U_{j,k} |^2 \lambda_k$ and canceled by $p_j$, which would occur if \smash{$\tr_{\shiftmath{-0.5pt}{\mbarsub}} (|\Phi _k '\rangle \langle \Phi _l '|)=\delta_{k,l}\tr_{\shiftmath{-0.5pt}{\mbarsub}} (|\Phi _k '\rangle \langle \Phi _k '|)=\delta_{k,l}\redxsuper{\rho}{m}{\prime}_{\text{MP}}$} where \smash{$\redxsuper{\rho}{m}{\prime}_{\shiftmath{-1pt}{\text{MP}}}$} is a state of MPSRP and needs to be independent of $k$. The $\delta_{k,l}$ and constant \smash{$\redxsuper{\rho}{m}{\prime}_{\text{MP}}$} let \Eq{21} cancel $U$ in \Eq{22}, achieving \Eq{24}, while \smash{$\tr_{\shiftmath{-0.5pt}{\mbarsub}} (|\Phi _k '\rangle \langle \Phi _k '|)=\redxsuper{\rho}{m}{\prime}_{\text{MP}}$} has MPSRP since the $|\Phi _k '\rangle$ are ME, satisfying \Eq{23}.
\item[\textbf{2.}]\hypertarget{Mechanism:2}{}\textit{However}, \smash{$\tr_{\mbarsub} (|\Phi _k '\rangle \langle \Phi _l '|)$} yielding a $\delta_{k,l}$ is \textit{not the only way} to achieve \Eq{24} \textit{and} \Eq{23}. Another way to achieve \Eq{24} (at least for one $m$) is if \smash{$\tr_{\mbarsub} (|\Phi _k '\rangle \langle \Phi _l '|)$} causes the right side of \Eq{22} to become a mixture with eigenstates each of which have the same coefficients from the same row $j$ of $U$ (allowing them to all be normalized by the same $p_j$), and whose eigenvalues yield a purity that is both MPSRP and independent of $U$, achieving \Eq{23}.
\end{itemize}
\vspace{-1pt}
In fact, for every mode $m$, \textit{either} of these mechanisms might be involved, but we only need to focus on one of them, as we explain next. (We \textit{could} drop \smash{$U_{\LU}$} here, but we keep it for generality and its role in the MSD.)

For each mode label $m$, there is an associated \textit{extreme bipartition} $(m|\mbar)$ of mode $m$ and all other modes (where $\mbar$ means ``not $m$'').  By the \textit{bipartite} Schmidt decomposition, \textit{the bigger-dimension part $B_m$ of $(m|\mbar)$ has the same spectrum as the smaller part $S_m$}, so we merely need to apply a version of \hyperlink{Mechanism:1}{Mechanism 1} to $S_m$ where \textit{the $S_m$-reduction} of $|\Phi _k '\rangle$ has the same purity as a mode-$m$ MPSRP state regardless of the dimension $n_{S_{m}}$ of $S_m$ for \Eq{23}.

In fact, in a \textit{bipartite} system of modes of different sizes, \hyperlink{Mechanism:1}{Mechanism 1} \textit{must} occur for the reduction to the \textit{smaller} mode $S_m$, since that is the only way that \Eq{23} and \Eq{24} can be satisfied for \textit{both} modes $m$.

For $N$-partite systems, the bipartite Schmidt decomposition still applies to each $(m|\mbar)$, but the MPSRP phenomenon also functions. For each $m$, either $m$ or $\mbar$ will be $S_m$. If $S_m =m$, \hyperlink{Mechanism:1}{Mechanism 1} guarantees \Eq{23} and \Eq{24} for that $m$ only, while the reduction purity of $B_m$ does not matter.  If $S_m =\mbar$, \hyperlink{Mechanism:1}{Mechanism 1} ensures that the index-matched operator\hsp{-1.5} \smash{$|\Phi _{\shiftmath{0.5pt}{k}} '\rangle \hsp{-1}\langle \Phi _{\shiftmath{0.5pt}{k}} '|$} is what gets traced over in \Eq{22} for \smash{$\redx{\rho}{\mbarsub}_{\shiftmath{-1.3pt}{|\Phi_k '\rangle}}$} [which is \smash{$\redx{\rho}{S_{\shiftmath{1pt}{m}}}_{\shiftmath{-1pt}{|\Phi_k '\rangle}}$}], while the fact that \smash{$|\Phi _k '\rangle \langle \Phi _k '|$} is ME means that the purity of \smash{$\redx{\rho}{\mbarsub}_{\shiftmath{-1.3pt}{|\hsp{0.8}\Phi_{\hsp{2.0}\shiftmath{1pt}{k}} '\rangle}}$}\hsp{-2.2} is whatever it needs to be for the purity of \smash{$\redx{\rho}{\hsp{0.3}m\hsp{0.3}}_{\shiftmath{-1.3pt}{|\hsp{0.8}\Phi_{\hsp{1.8}k} '\rangle}}$}\hsp{-2} to be an MPSRP state given the bipartite\hsp{-1} Schmidt\hsp{-1} decomposition of $(m|\mbar)$. Then if \smash{$\redx{\rho}{\mbarsub}_{\shiftmath{-1.3pt}{|\Phi_k '\rangle}}$} is $k$-independent, \Eq{23} is achieved.

Thus, candidate sufficient MME conditions in terms of \textit{smaller} parts $S_m$ and a modified \hyperlink{Mechanism:1}{Mechanism 1} are
\begin{Equation}                      {30}
\tr_{\overline{S_m} } (|\Phi _k '\rangle \langle \Phi _l '|) = \delta _{k,l} \redx{\rho}{S_{m}}_{|\Phi _k '\rangle};\;\; \left\{\! {\begin{array}{*{20}l}
   {k,l \in 1, \ldots ,r;}  \\
   {\redx{\rho}{S_{m}}_{|\Phi _k '\rangle}=\redx{\rho}{S_{m}}_{|\Phi '\rangle}\;\forall k;} \\
   {P(\redx{\rho}{S_{m}}_{|\Phi '\rangle})=P(\redxsuper{\rho}{m}{\prime}_{\text{MP}});}  \\
   {\forall m \in 1, \ldots ,N,}  \\
\end{array}} \right.
\end{Equation}
where \smash{$\redx{\rho}{S_{m}}_{|\Phi _k '\rangle}$} is the mode-$S_m$ reduction of \smash{$|\Phi _k '\rangle$}, $S_m$ is the mode-group label of the smaller extreme bipartition of which mode $m$ is a member (so if $n_{S_m }  = \min \{ n_m ,n_{\mbarsub} \}  = n_m$, then $S_m  = m$, but if $n_{S_m }  = \min \{ n_m ,n_{\mbarsub} \}  = n_{\mbarsub}$, then $S_m  = \mbar$), where again $\mbar$ is a list of all modes \textit{except} $m$ (thus, \smash{$\overline{S_m}=B_m$}). Note that \Eq{30} requires the $S_{m}$ reduction of all \smash{$|\Phi _k '\rangle$} to be identical to \smash{$\redx{\rho}{S_{m}}_{\shiftmath{-1.3pt}{|\Phi '\rangle}}$} (for some ME state $|\Phi '\rangle$)\hsp{-0.9} and\hsp{-0.9} thus\hsp{-0.9} independent\hsp{-0.9} of\hsp{-0.9} $k$\hsp{-0.9} for\hsp{-0.9} each\hsp{-0.9} $m$.

To see if the candidate conditions of \Eq{30} are truly sufficient, we need to show that \Eq{30} leads to MME states (in fact there are a few missing ingredients which we will motivate below and then include). So, putting \Eq{29} into \Eq{30} for the case where $S_m =m$ gives, $\forall m \in 1, \ldots ,N$,
\begin{Equation}                      {31}
\begin{array}{*{20}l}
   {\tr_{\overline{S_m} } (|\Phi _k '\rangle \langle \Phi _l '|)} &\!\! {=\tr_{\mbarsub } (|\Phi _k '\rangle \langle \Phi _l '|)}  \\
   {} &\!\! {=\frac{1}{{L_* }}\sum\limits_{b = 1}^{L_* } {} \sum\limits_{c = 1}^{L_* } {} \tr\left( {|U_{:,v_{k,b}^{(1)} }^{(1)} \rangle \langle U_{:,v_{l,c}^{(1)} }^{(1)} |} \right) \otimes  \cdots }  \\
   {} &\!\! {\hsp{11} \otimes \tr\left( {|U_{:,v_{k,b}^{(m  - 1)} }^{(m  - 1)} \rangle \langle U_{:,v_{l,c}^{(m  - 1)} }^{(m  - 1)} |} \right)}  \\
   {} &\!\! {\hsp{11} \otimes |U_{:,v_{k,b }^{(m)} }^{(m )} \rangle \langle U_{:,v_{l,c }^{(m )} }^{(m )} |}  \\
   {} &\!\! {\hsp{11} \otimes \tr\left( {|U_{:,v_{k,b}^{(m  + 1)} }^{(m  + 1)} \rangle \langle U_{:,v_{l,c}^{(m  + 1)} }^{(m  + 1)} |} \right) \otimes  \cdots }  \\
   {} &\!\! {\hsp{11} \otimes \tr\left( {|U_{:,v_{k,b}^{(N)} }^{(N)} \rangle \langle U_{:,v_{l,c}^{(N)} }^{(N)} |} \right)\!,}  \\
\end{array}
\end{Equation}
while for the $S_m =\mbar$ case it gives, $\forall m \in 1, \ldots ,N$,
\begin{Equation}                      {32}
\begin{array}{*{20}l}
   {\tr_{\overline{S_m} } (|\Phi _k '\rangle \langle \Phi _l '|)} &\!\! {=\tr_{m } (|\Phi _k '\rangle \langle \Phi _l '|)} \\
   {} &\!\! {=\frac{1}{{L_* }}\sum\limits_{b = 1}^{L_* } {} \sum\limits_{c = 1}^{L_* } {}  {|U_{:,v_{k,b}^{(1)} }^{(1)} \rangle \langle U_{:,v_{l,c}^{(1)} }^{(1)} |}  \otimes  \cdots }  \\
   {} &\!\! {\hsp{11} \otimes {|U_{:,v_{k,b}^{(m  - 1)} }^{(m  - 1)} \rangle \langle U_{:,v_{l,c}^{(m  - 1)} }^{(m  - 1)} |} }  \\
   {} &\!\! {\hsp{11} \otimes \tr\left({|U_{:,v_{k,b }^{(m)} }^{(m )} \rangle \langle U_{:,v_{l,c }^{(m )} }^{(m )} |}\right)}  \\
   {} &\!\! {\hsp{11} \otimes {|U_{:,v_{k,b}^{(m  + 1)} }^{(m  + 1)} \rangle \langle U_{:,v_{l,c}^{(m  + 1)} }^{(m  + 1)} |}  \otimes  \cdots }  \\
   {} &\!\! {\hsp{11} \otimes  {|U_{:,v_{k,b}^{(N)} }^{(N)} \rangle \langle U_{:,v_{l,c}^{(N)} }^{(N)} |}, }  \\
\end{array}
\end{Equation}

Then, using \smash{$|U_{\shiftmath{1.2pt}{\hsp{4}:,\hsp{2.0}v_{j,h}^{(m)}}}^{(m)} \rangle\,\equiv\, U^{(m)}|v_{j,h}^{(m)} \rangle$} [from the text after \Eq{15}],\hsp{0.8} followed\hsp{0.8} by\hsp{1} the unitary invariance of the trace $\tr(UAU^{\dag})=\tr(AU^{\dag}U)=\tr(A)$, \Eq{31} simplifies to
\begin{Equation}                      {33}
\begin{array}{*{20}l}
   {\tr_{\overline{S_m} } (|\Phi _k '\rangle \langle \Phi _l '|)} &\!\! {=\tr_{\mbarsub } (|\Phi _k '\rangle \langle \Phi _l '|)} \\
   {} &\!\! {=\frac{1}{{L_* }}\sum\limits_{b = 1}^{L_* } {} \sum\limits_{c = 1}^{L_* } {} \delta _{v_{k,b}^{(1)} ,v_{l,c}^{(1)} }  \cdots \delta _{v_{k,b}^{(m  - 1)} ,v_{l,c}^{(m  - 1)} } }  \\
   {} &\!\! {\phantom{=} \times U^{(m)} |v_{k,b }^{(m )} \rangle \langle v_{l,c }^{(m )} |U^{(m)\dag}}  \\
   {} &\!\! {\phantom{=} \times \delta _{v_{k,b}^{(m  + 1)} ,v_{l,c}^{(m  + 1)} }  \cdots \delta _{v_{k,b}^{(N)} ,v_{l,c}^{(N)} } ,}  \\
\end{array}\!\!\!
\end{Equation}
$\forall m \in 1, \ldots ,N$, while \Eq{32} simplifies to
\begin{Equation}                      {34}
\begin{array}{*{20}l}
   {\tr_{\overline{S_m} } (|\Phi _k '\rangle \langle \Phi _l '|)} &\!\!\! {=\!\tr_{m } (|\Phi _k '\rangle \langle \Phi _l '|)} \\
   {} &\!\!\! {=\!\frac{1}{{L_* }}\sum\limits_{b = 1}^{L_* } {} \sum\limits_{c = 1}^{L_* } {}  U^{(1)}|v_{k,b}^{(1)} \rangle {\kern -1pt}\langle v_{l,c}^{(1)} |U^{(1)\dag} \! \otimes  \cdots }  \\
   {} &\!\!\! {\phantom{=}\! \otimes U^{(m  - 1)}|v_{k,b}^{(m  - 1)}\rangle {\kern -1pt}\langle v_{l,c}^{(m  - 1)} |U^{(m  - 1)\dag} }  \\
   {} &\!\!\! {\phantom{=}\! \times \delta_{v_{k,b }^{(m)},v_{l,c }^{(m )}} }  \\
   {} &\!\!\! {\phantom{=}\! \otimes U^{(m  +\! 1)}|v_{k,b}^{(m  +\! 1)} \rangle {\kern -1pt}\langle v_{l,c}^{(m  +\! 1)} |U^{(m  +\! 1)\dag}  \!\otimes\!  \cdots }  \\
   {} &\!\!\! {\phantom{=}\! \otimes  U^{(N)}|v_{k,b}^{(N)} \rangle{\kern -1pt}\langle v_{l,c}^{(N)} |U^{(N)\dag}, }  \\
\end{array}
\end{Equation}
$\forall m \in 1, \ldots ,N$. Notice that the single-mode unitaries $U^{(m)}$ of \Eq{29} being traced over in \Eq{31} and \Eq{32} have no effect on \Eq{33} and \Eq{34}, and those that remain in $S_m$ will have no effect on the purity of \smash{$\redx{\rho}{S_{m}}_{|\Phi _k '\rangle}$}.

In \Eq{33} and \Eq{34} it is useful to recall that, given some arbitrarily numerically labeled set of ME TGX tuples, \smash{$\mathbf{v}_{k,b}$} is the coincidence vector index of the $b$th level of the $k$th ME TGX tuple, and that its $m$th component \smash{$v_{k,b}^{(m)}$} corresponds to a computational basis state of mode $m$. Therefore, for \Eq{33} and \Eq{34} to achieve the  $\delta _{k,l}$ specified by \Eq{30}, consider the following facts. 
\begin{itemize}[leftmargin=*,labelindent=4pt]\setlength\itemsep{0pt}
\item[\textbf{1.}]\hypertarget{Fact:1}{}The levels \textit{within} any $k$th ME TGX tuple are always orthogonal, so when $k=l$, we are guaranteed that \smash{$\delta _{\shiftmath{1.4pt}{v_{k,b}^{(m)},\!v_{l,c}^{(m)} }}=\delta _{\shiftmath{1.4pt}{v_{k,b}^{(m)},\!v_{k,c}^{(m)} }}=\delta_{b,c}$} \textit{for at least one $m$} (since in the coincidence basis, only one mode needs a different index for any two outcomes to be orthogonal).
\item[\textbf{2.}]\hypertarget{Fact:2}{}On the other hand, $k\neq l$ implies two \textit{different} ME TGX tuples. Since the deltas in \Eq{33} and \Eq{34} \textit{do not rely on phases} but still need to yield $\delta_{k,l}$ in either $S_m$ case, and since the indices \smash{$v_{\shiftmath{0.8pt}{k,b}}^{\shiftmath{2pt}{(m)}}$} are vector-index components of levels of ME TGX tuples of the core ME TGX states \smash{$\{|\Phi _k \rangle\}$}, which by \Eq{29} must be orthonormal, then a new condition to add to \Eq{30} is that\\
\vspace{-21pt}
\begin{Equation}                      {35}
|\Phi _k \rangle\;\text{must be \textit{phaseless} for }k\in 1,\ldots,r,
\end{Equation}
\vspace{-14pt}\\
meaning their nonzero coefficients are only real and positive. Note that in the MEB theorem, the ME TGX states generally need phases, but there is always a subset of them that can be chosen to be phaseless.  Furthermore, since these $|\Phi _k \rangle$ must be orthonormal \textit{and} have no phases to enable that, then another condition to add to \Eq{30} is that
\vspace{-1pt}
\begin{Equation}                      {36}
\left( {\begin{array}{*{20}c}
   {\parbox{2.4in}{the phaseless core ME TGX states must be \textit{spacewise orthogonal} $\{|\Phi _k \rangle\}\in\oplus^{\perp}$}}  \\
\end{array}} \right)\!,
\end{Equation}
\vspace{-8pt}\\
where states are \textit{spacewise orthogonal} \smash{$\oplus^{\perp}$} iff they reside in orthogonal subspaces. Thus, MME-compatible ME TGX tuples have no matching levels in scalar-index form [Alternatively, \Eq{36} is equivalent to requiring spacewise orthogonality of each $B_m$ reduction of $\{|\Phi _k \rangle\}$]. Thus, \Eq{36} guarantees that if $k\neq l$, then {$\delta _{\shiftmath{1.4pt}{v_{k,b}^{(m)},\!v_{l,c}^{(m)}} }=0$} \textit{for at least one $m$}, \textit{provided that such ME TGX states exist}, which leads us to the next point.
\item[\textbf{3.}]\hypertarget{Fact:3}{}\Eq{36} implies another condition;
\vspace{-1pt}
\begin{Equation}                      {37}
\left( {\begin{array}{*{20}c}
   {\parbox{2.0in}{the $r$ phaseless ME TGX states $|\Phi _k \rangle$ needed to achieve \Eq{30} need to be able to \textit{exist} in the system}}  \\
\end{array}} \right)\!,
\end{Equation}
\vspace{-8pt}\\
since, from \cite{HedE}, there are only a finite number of ME TGX tuples in any finite discrete system, and each phaseless ME TGX state has a unique ME TGX tuple.
\item[\textbf{4.}]\hypertarget{Fact:4}{}Thus, given \hyperlink{Fact:1}{Facts 1--3}, if \Eqs{35}{37} are taken as refinements to \Eq{30}, if they are fulfilled, then
\begin{Equation}                      {38}
\left\{\!\!\! {\begin{array}{cll}
   {\left(\!\!\! {\begin{array}{l}
   {\phantom{\times}\delta _{v_{k,b}^{(1)} ,v_{l,c}^{(1)} }  \cdots \delta _{v_{k,b}^{(m - 1)} ,v_{l,c}^{(m - 1)} } }  \\
   { \times \delta _{v_{k,b}^{(m + 1)} ,v_{l,c}^{(m + 1)} }  \cdots \delta _{v_{k,b}^{(N)} ,v_{k,b}^{(N)} } }  \\
\end{array}}\!\! \right)} &\!\!\! { = \delta _{k,l} \delta _{b,c} ;} & {S_m  \!=\! m}  \\
   {\delta _{v_{k,b}^{(m)} ,v_{l,c}^{(m)} } } &\!\!\! { = \delta _{k,l} \delta _{b,c} ;} & {S_m  \!=\! \mbar.}  \\
\end{array}} \right.
\end{Equation}
\end{itemize}
Thus, adopting \Eqs{35}{37} as refinements to \Eq{30}, then given \hyperlink{Fact:1}{Facts 1--4}, if a set of $r$ spacewise orthogonal ME TGX tuples exists in a system, the deltas in \Eq{33} and \Eq{34} collectively yield $\delta_{k,l}\delta_{b,c}$, so \Eq{33} and \Eq{34} simplify to
\begin{Equation}                      {39}
\begin{array}{*{20}l}
   {\tr_{\overline{S_m} } (|\Phi _k '\rangle \langle \Phi _l '|)=\delta_{k,l}\tr_{\overline{S_m} } (|\Phi _k '\rangle \langle \Phi _k '|) }  \\
   {\quad =\! \left\{\!\! {\begin{array}{*{20}l}
   {\delta_{k,l}U^{(m)}\!\left({\frac{1}{{L_* }}\!\sum\limits_{b = 1}^{L_* } {}\! |v_{k,b }^{(m )} \rangle \langle v_{k,b }^{(m )} |}\right)\!U^{(m)\dag};} & {S_{m}\!=\!m}  \\
   {\delta_{k,l}U^{(\mbarsub)}\!\left({\frac{1}{{L_* }}\!\sum\limits_{b = 1}^{L_* } {}\! |\mathbf{v}_{k,b }^{(\mbarsub )} \rangle \langle \mathbf{v}_{k,b }^{(\mbarsub )} |}\right)\!U^{(\mbarsub)\dag};} & {S_{m}\!=\!\mbar,}  \\
\end{array}} \right.}  \\
\end{array}\!
\end{Equation}
where \smash{$U^{(\mbarsub)}\!\equiv\! U^{(1)}\!\otimes\cdots\otimes\! U^{(m-1)}\!\otimes\! U^{(m+1)}\!\otimes\cdots\otimes\! U^{(N)}$} and \smash{$\!|\mathbf{v}_{k,b }^{(\mbarsub )} \rangle\equiv|v_{k,b }^{(1)} \rangle\cdots|v_{k,b }^{(m-1 )} \rangle|v_{k,b }^{(m+1 )} \rangle\cdots|v_{k,b }^{(N)} \rangle$}\rule{0pt}{12pt}.

To achieve the requirement of \Eq{30} that all $S_m$\rule{0pt}{13pt} reductions be the same, the diagonal states on the right in \Eq{39} need to either be full-rank in $S_m$ or less-than-full-rank but have identical elements for all \smash{$\redx{\rho}{S_{m}}_{|\Phi _k '\rangle}$}.

From App.D.4.c of \cite{HedE},\rule{0pt}{13pt} two general cases emerge for full $N$-partite entanglement; Case 1: systems with multiple largest modes (Possibilities 1 and 3 in \cite{HedE}) and Case 2: systems with a single largest mode (Possibility 2 in \cite{HedE}). In Case 1, $S_m  = m$ is the only case that applies for all extreme bipartitions $(m|\mbar)$ and since ME $N$-partite states of those systems have ideal maximal mixing in every mode-$m$ reduction, those states are \textit{also ME for each} $(m|\mbar)$, which guarantees ME $N$-partite states have maximally mixed $S_m$ reductions.  Case 2 acts like Case 1 except for the largest mode $m_{\max}$, for which examples show that either $S_{m_{\max}}  = m_{\max}$ or $S_{m_{\max}}  = \overline{m_{\max}\rule{0pt}{5pt}}$ can happen, and in both subcases, achieving maximally mixed $S_{m_{\max}}$ reductions depends on $n_{S_{m_{\max}}}$ relative to $L_*$. Thus,
\begin{Equation}                      {40}
\left\{ \! {\begin{array}{*{20}l}
   {\left(\; {\parbox{2.0in}{If $\mathbf{n}$ has multiple largest modes, then \smash{$\redx{\rho}{S_{m}}_{|\Phi\rangle}=\frac{1}{n_{S_m}}I^{(S_{m})}\; \forall m$}.}}\; \right)}  \\
   {\!\left(\,\, {\parbox{2.7in}{If $\mathbf{n}$ has only one largest mode $m_{\max}$, then \smash{$\redx{\rho}{S_{m}}_{|\Phi\rangle}\hsp{-0.5}=\hsp{-0.5}\frac{1}{n_{S_m}}I^{(S_{m})}\;\,\forall m\neq m_{\max}$},\rule{0pt}{10pt} but achieving \smash{$\redx{\rho}{S_{m_{\max}}}_{\hsp{-0.7}\shiftmath{-1pt}{|\Phi\rangle}}=\shiftmath{1pt}{\frac{1}{n_{S_{m_{\max}}}}}I^{(S_{m_{\max}})}$}\rule{0pt}{11pt} depends on the\hsp{0.9} value\hsp{0.8} of\hsp{0.8} \smash{$n_{S_{m_{\max}}}$}\hsp{0.8} relative\hsp{0.8} to\hsp{0.8} $L_*$\rule{0pt}{10pt}.}}\,\, \right)\!,\rule{0pt}{37pt}}  \\
\end{array}} \right.
\end{Equation}
where $m_{\max}$ is the mode of dimension $n_{\max} \equiv n_{m_{\max}}\equiv\max\{\mathbf{n}\}$ and $|\Phi\rangle$ is any $N$-partite ME TGX state.

Thus, \Eq{40} and the form of \Eq{39} show that we may be able to fulfill the $k$-independence requirement of \Eq{30}, but in Case 2, if \smash{$\redx{\rho}{S_{m_{\max}}}_{|\hsp{1}\Phi_{\hsp{1}k} '\rangle}$} are not full-rank, they need to be identical $\forall k$, \textit{and} are subject to ME TGX tuple availability by \Eq{37} due to \Eq{29}. Therefore, that requirement of \Eq{30} \textit{still} needs to be specified as
\begin{Equation}                      {41}
\{|\Phi _k '\rangle\} \;\;\text{need to have}\;\;\, \redx{\rho}{S_{m}}_{|\Phi _k '\rangle}=\redx{\rho}{S_{m}}_{|\Phi '\rangle}\;\;\forall k\;\forall m.
\end{Equation}
Then, since \smash{$\tr_{\overline{S_m} } (|\Phi _k '\rangle \langle \Phi _k '|)=\redx{\rho}{S_{\shiftmath{1pt}{m}}}_{\shiftmath{-1.1pt}{|\hsp{1}\Phi _{\hsp{1}\shiftmath{1pt}{k}} '\rangle}}$} in \Eq{39}, \textit{supposing that\hsp{-0.5} it\hsp{-0.5} satisfies\hsp{-0.5} \Eq{41}\hsp{-0.5} and\hsp{-0.5} \Eq{37}},\hsp{-0.5} putting \Eq{41} into \Eq{39} gives
\begin{Equation}                      {42}
\tr_{\overline{S_m} } (|\Phi _k '\rangle \langle \Phi _l '|) =\left\{\!\! {\begin{array}{*{20}l}
   {\delta _{k,l}\redx{\rho}{m}_{|\Phi '\rangle}  \!=\! \delta _{k,l} \redxsuper{\rho}{m}{\prime}_{\text{MP}} ;} & {S_m  \!=\! m}  \\
   {\delta _{k,l}\redx{\rho}{\mbarsub}_{|\Phi '\rangle};} & {S_m  \!=\! \mbar.}  \\
\end{array}} \right.
\end{Equation}
\vspace{-18pt}\\

Then, in the $S_m  \!=\! m$ case, putting \Eq{42} into \Eq{22} and using \Eq{21} yields, $\forall j$,
\begin{Equation}                      {43}
\begin{array}{*{20}l}
   {\redx{\rho}{m}_{|w_{j}\rangle}} &\!\! {=\frac{1}{{p_j }}\sum\limits_{k = 1}^{r} {} \sum\limits_{l = 1}^{r} {} U_{j,k} U_{j,l} ^* \sqrt {\lambda_k \lambda_l } \delta _{k,l} \redxsuper{\rho}{m}{\prime}_{\text{MP}}}  \\
   {} &\!\! {=\redxsuper{\rho}{m}{\prime}_{\text{MP}}\frac{1}{{p_j }}\sum\limits_{k = 1}^{r} {} |U_{j,k}|^2 \lambda_k  =\redxsuper{\rho}{m}{\prime}_{\text{MP}}.}  \\
\end{array}
\end{Equation}
In the $S_m  \!=\! \mbar$ case, looking at the reduction for mode group $\mbar$ (which is generally not one of the single modes of the system except for bipartite systems), we get, $\forall j$,
\begin{Equation}                      {44}
\begin{array}{*{20}l}
   {\redx{\rho}{\mbarsub}_{|w_{j}\rangle}} &\!\! {=\frac{1}{{p_j }}\sum\limits_{k = 1}^{r} {} \sum\limits_{l = 1}^{r} {} U_{j,k} U_{j,l} ^* \sqrt {\lambda_k \lambda_l } \delta _{k,l}\redx{\rho}{\mbarsub}_{|\Phi '\rangle}}  \\
   {} &\!\! {=\redx{\rho}{\mbarsub}_{|\Phi '\rangle}\frac{1}{{p_j }}\sum\limits_{k = 1}^{r} {} |U_{j,k}|^2 \lambda_k  =\redx{\rho}{\mbarsub}_{|\Phi '\rangle},}  \\
\end{array}
\end{Equation}
where the fact that \smash{$\shiftmath{1pt}{\redx{\rho}{\mbarsub}_{\hsp{-1}\shiftmath{-1pt}{|w_{j}\rangle}}}$} is the \textit{smaller} reduction of a pure state $|w_{j}\rangle$ over $(m|\mbar)$ means that the bipartite Schmidt decomposition over $(m|\mbar)$ causes \smash{$\redx{\rho}{m}_{\hsp{-1}\shiftmath{-1pt}{|w_{j}\rangle}}$} to have the \textit{same spectrum} as \smash{$\redx{\rho}{\mbarsub}_{\hsp{-1}\shiftmath{-1pt}{|w_{j}\rangle}}$}, so their purities are equal as
\begin{Equation}                      {45}
P(\redx{\rho}{m}_{|w_{j}\rangle})=P(\redx{\rho}{\mbarsub}_{|w_{j}\rangle}).
\end{Equation}
Then, putting \Eq{44} into \Eq{45} gives
\vspace{-4pt}
\begin{Equation}                      {46}
P(\redx{\rho}{m}_{|w_{j}\rangle})=P(\redx{\rho}{\mbarsub}_{|\Phi '\rangle});\;\;(S_m =\mbar),
\end{Equation}
\vspace{-12pt}\\
which is \textit{almost} what we need, but due to Case 2 in \Eq{40}, not all systems have $N$-partite ME states $|\Phi '\rangle$ that are also ME for systems of all of their extreme bipartitions $(m|\mbar)$. Therefore, for the case of $S_m  = \mbar$, we also \textit{still} need the next condition of \Eq{30},
\vspace{-4pt}
\begin{Equation}                      {47}
\text{ME states}\;\;|\Phi '\rangle\;\;\text{need to have}\;\; P(\redx{\rho}{\mbarsub}_{|\Phi '\rangle})=P(\redxsuper{\rho}{m}{\prime}_{\text{MP}}).
\end{Equation}
Therefore, \textit{if \Eq{47} is also fulfilled}, then \Eq{46} becomes
\begin{Equation}                      {48}
P(\redx{\rho}{m}_{|w_{j}\rangle})=P(\redxsuper{\rho}{m}{\prime}_{\text{MP}});\;\;(S_m =\mbar).
\end{Equation}
Since \Eq{48} implies that \smash{$\redx{\rho}{m}_{|w_{j}\rangle}$} is an MPSRP \textit{state}, but this state does \textit{not} have to\hsp{-1} be constant $\forall j$ (since $m\neq S_m$) except for its purity [which \textit{is} constant by \Eq{48}], then
\begin{Equation}                      {49}
\redx{\rho}{m}_{|w_{j}\rangle}\!=\!\redxsuper{\rho}{m}{\prime}_{\text{MP}_{j}}\;\;\text{s.t.}\;\;P(\redxsuper{\rho}{m}{\prime}_{\text{MP}_{j}})\!=\!P(\redxsuper{\rho}{m}{\prime}_{\text{MP}})\;\;\forall j;\;\;(S_m =\mbar),
\end{Equation}
where the $j$ in \smash{$\redxsuper{\rho}{m}{\prime}_{\text{MP}_{j}}$} lets \smash{$\{\shiftmath{1.5pt}{\redx{\rho}{m}_{\hsp{-1}\shiftmath{-1pt}{|w_{j}\rangle}}}\}$} be different nonfull-rank MPSRP states wrt $j$, but with constant purities $\forall j$.

Since \Eq{43} and \Eq{49} \textit{also} both have $U$-independent purities (since \smash{$\redxsuper{\rho}{m}{\prime}_{\text{MP}}$} are reductions of the $U$-independent $|\Phi _k '\rangle$ states), then together they yield (uniting $S_m$ cases),
\begin{Equation}                      {50}
\redx{\rho}{m}_{|w_{j}\rangle} \! =\! \redxsuper{\rho}{m}{\prime}_{\text{MP}_{j}}\;\;\text{s.t.}\;\;P(\redxsuper{\rho}{m}{\prime}_{\text{MP}_{j}})\!=\!P(\redxsuper{\rho}{m}{\prime}_{\text{MP}}) ;\;\;\;\forall m,\;\forall j,\;\forall U,
\end{Equation}
and since \Eq{50} achieves \Eq{23} and \Eq{24}, \textit{our set of sufficient conditions of \Eq{30} and \Eqs{35}{37} achieves the necessary multipartite MME conditions}. 

However, since \Eq{30} and \Eqs{35}{37} contain some constraints that guarantee each other, they can be simplified a bit.  First, since achieving \Eq{50} would fail if \Eq{41} were not achieved, then the only systems with MME potential are those with sets of ME states $\{|\Phi _k '\rangle\}$ that satisfy \Eq{41}, and since all ME states $|\Phi _k '\rangle$ already obey \Eq{50}, then any set $\{|\Phi _k '\rangle\}$ that achieves \Eq{41} is already guaranteed to produce the $S_m$ reductions of \Eq{30} satisfying all side constraints. Therefore, we can trace over \Eq{30} without changing those facts. Also, since the LU in $|\Phi _k '\rangle=U_{\LU}|\Phi _k \rangle$ has no effect on entanglement since it cannot affect reduction purities, we can ignore it and simply write all conditions in terms of ME TGX states $|\Phi _k \rangle$. Therefore, also using \smash{$\overline{S_m}=B_m$}, \Eq{30} can be rewritten without changing its meaning as
\vspace{-2pt}
\begin{Equation}                      {51}
\tr[\tr_{B_m } (|\Phi _k \rangle \langle \Phi _l |)] = \delta _{k,l} ;\quad \left\{ {\begin{array}{*{20}l}
   {k,l \in 1, \ldots ,r;}  \\
   {\forall m \in 1, \ldots ,N.}  \\
\end{array}} \right.
\end{Equation}
\vspace{-20pt}\\

All of this means that for a given $L_*$ and rank $r$, the existence of MME in any system requires that \textit{each} $B_m$ is at least as big as the lowest number of pure decomposition states $r$ times the number of levels $L_*$ in the ME TGX eigenstates.  So, for given $L_*$ and $r$, the MME conditions in \Eq{51} imply a generally \textit{loose} upper bound of
\vspace{-4pt}
\begin{Equation}                      {52}
\min\{n_{B_1 },\ldots,n_{B_N }\}\geq rL_* ,
\end{Equation}
where \smash{$n_{B_m }  \equiv \max \{ n_m ,n_{\mbarsub} \}$} for each $m\in 1,\ldots,N$, since if the \textit{smallest} $n_{B_m}$ is at least $rL_*$, all other $n_{B_m}$ will be as well. The reason this is not always a tight upper bound is that \textit{\Eq{52} is constrained by the availability of ME TGX tuples in the system that satisfy \Eq{51}}, due to \Eq{37}. To account for this, let \smash{$\{|\Phi_{\ME_{\TGX}}^{\specialstar}\rangle\}$} be the set of ME TGX states that satisfy both \Eq{51} and \Eq{52}, and amend \Eq{52} to
\begin{Equation}                      {53}
\min\{\mathbf{n}_{B}\}\geq rL_* ;\;\;\text{s.t.}\;\; \exists\{|\Phi_{\ME_{\TGX}}^{\specialstar}\rangle\},
\end{Equation}
where $\mathbf{n}_{B}\equiv(n_{B_1 },\ldots,n_{B_N })$. \textit{Technically}, since \Eq{53} is valid for \textit{any} $L_* \in \mathbf{L}_*$, we could have TGX eigenstates of multiple \textit{different} $L_*$ values, so $r$ would be determined \textit{implicitly} by \smash{$\min\{\mathbf{n}_{B}\}\geq \sum\nolimits_{q=1}^{r} L_{*q}'$} \smash{$\text{s.t.}\; \exists\{|\Phi_{\ME_{\TGX}}^{\specialstar}\rangle\}$}, where\hsp{-0.5} $L_{*q}'$\hsp{-0.5} is\hsp{-0.5} one\hsp{-0.5} of\hsp{-0.5} the\hsp{-0.5} $L_*$\hsp{-0.5} in\hsp{-0.5} $\mathbf{L}_*$, where the prime means we allow repeated values. However, by \Eq{29}, \textit{single}-$L_*$ states will suffice. Thus to achieve \Eq{25}, to find the \textit{largest MME rank} $r$, we merely need to \textit{minimize} $L_*$ in \Eq{53}, as
\begin{Equation}                      {54}
\min\{\mathbf{n}_{B}\}\geq r \min\{\mathbf{L}_*\} ;\;\;\text{s.t.}\;\; \exists\{|\Phi_{\ME_{\TGX}}^{\specialstar}\rangle\},
\end{Equation}
or equivalently, we can get all possible MME ranks as
\begin{Equation}                      {55}
2\leq r\leq\text{floor}\left({\frac{\min\{\mathbf{n}_{B}\}}{\min\{\mathbf{L}_*\}}}\right)\! ; \;\;\text{s.t.}\;\; \exists\{|\Phi_{\ME_{\TGX}}^{\specialstar}\rangle\}.
\end{Equation}

In \textit{bipartite} systems, \Eq{54} simplifies to $n_B  \geq rL_* $ for $r\geq 2$, and since bipartite systems always have $L_* = n_S$, \Eq{54} simplifies further to $n_B  \geq rn_S $ as given before \Eq{3} (subscript $m$ is not needed since in a $2$-mode system, picking either mode as the single mode $m$ to define the extreme bipartition yields the same extreme bipartition).

Thus, by \Eq{55} the largest rank that supports MME is
\begin{Equation}                      {56}
\maxMMErank  \equiv\! \mathop {\max \{2,\ldots,\maxMMEranklim\} ; }\limits_{\scalemath{0.98}{\begin{array}{*{20}l}
   {\text{s.t.}\; \exists\{|\Phi_{\ME_{\TGX}}^{\specialstar}\rangle\}}  \\
\end{array}}} \;\;\;{{\scriptstyle \maxMMEranklim \,\equiv\, \text{floor}}\!\left(\! {\textstyle \frac{{\min \{ \mathbf{n}_B \} }}{{\min \{ \mathbf{L}_* \} }}}\!\right)\!,}
\end{Equation}
where $\maxMMEranklim$ is the \textit{loose maximal MME rank upper limit}. To more clearly express the fact that \smash{$\{|\Phi_{\ME_{\shiftmath{1pt}{\TGX}}}^{\specialstar}\rangle\}$} implies that the maximization in \Eq{56} is constrained by a search over \textit{all} ME TGX tuples \smash{$\{|\Phi _{\ME_{\TGX} } \rangle\}$} subject to \Eq{51}, we replace the maximization argument by rank $r$ bounded by its loosest limits (reducing the lower limit to $1$ to include pure states as a trivial case), and specify the constraints of \Eq{51} beneath it to get the \textit{maximal MME rank} as 
\begin{Equation}                      {57}
\maxMMErank  \equiv\!\!\! \mathop {\max \{ r\}_1^\maxMMEranklim ; }\limits_{\scalemath{0.95}{\begin{array}{*{20}l}
   {\text{s.t.}\,\,\exists \{ |\Phi _k \rangle \}_{k = 1}^{r}\!\in\!\{|\Phi _{\ME_{\TGX} } \rangle\};}  \\
   {\tr[\tr_{B_m } (|\Phi _k \rangle \langle \Phi _l |)]\! =\! \delta _{k,l}\,\,\forall m }  \\
\end{array}}} {\kern -6pt}{{\scriptstyle \maxMMEranklim \,\equiv\, \text{floor}}\!\left(\! {\textstyle \frac{{\min \{ \mathbf{n}_B \} }}{{\min \{ \mathbf{L}_* \} }}}\!\right)}
\end{Equation}
which gives \Eq{5}, where \smash{$\{r\}_{1}^{\maxMMEranklim}$} means \smash{$r\in 1,\ldots,\maxMMEranklim$}, $\{ |\Phi _k \rangle \}_{k = 1}^{r}\equiv \{|\Phi _1 \rangle,\ldots,|\Phi _r \rangle\}$, and $m\in 1,\ldots,N$.

Thus, to see if a system can host MME, just use \Eq{57}; if $\maxMMErank\geq 2$, it can host MME for ranks $r\in 2,\ldots, \maxMMErank$, whereas $\maxMMErank=1$ means \textit{strict} MME is not possible in that system, and instead corresponds to the trivial mixture where $\rho$ is a \textit{pure} ME state (in the pure-state-inclusive definition of mixtures). Thus, $\maxMMErank$ is a transparent and informative indicator of MME potential.

Next, we show how to apply \Eq{57} with examples, and test its predictions by  demonstrating their validity in a variety of multipartite systems.
\section{\label{sec:V}Tests and Examples of Multipartite MME}
Since bipartite ($N = 2$) MME has already been extensively covered in \cite{LZFF}, here we focus mainly on systems with $N\geq 3$ modes. However, for thoroughness, we present tables and examples for $N\geq 2$.

Here, we start with a bipartite example of how to identify TGX eigenstates for an MME state in $2\times 5$, which will help make the $2\times 2\times 2\times 2$ example after it easier to understand. These examples will also help show how to implement \Eq{5} in a practical way.
\subsection{\label{sec:V.A}Example for Bipartite MME}
In $2\times 5$, to construct an MME state, first use the 13-step algorithm $\mathcal{A}_{13}$ of \cite{HedE} to generate all the ME TGX tuples for each starting level $S_L$ (the level that is guaranteed to be in all ME TGX tuples generated by $\mathcal{A}_{13}$ with $S_L$ as input).  For $S_L =1$, the ME TGX tuples are
\begin{Equation}                      {58}
\{ 1,7\} ,\{ 1,8\} ,\{ 1,9\} ,\{ 1,10\} ;\;\;\text{choose}\;\,\{ 1,10\} .
\end{Equation}
Since all ME TGX eigenstates of an MME state must be spacewise orthogonal by \Eq{36}, we can only choose \textit{one} of these, so suppose we choose $\{ 1,10\}$. Continuing our search, the $S_L =2$ ME TGX tuples are
\begin{Equation}                      {59}
\{ 2,6\} ,\{ 2,8\} ,\{ 2,9\} ,\{ 2,10\} ;\;\;\text{choose}\;\,\{ 2,8\},
\end{Equation}
and since we already chose $\{1,10\}$ as the first pair, the spacewise orthogonality requirement  means we can choose any tuple in this set \textit{except} $\{ 2,10\}$ since it includes level $10$, so suppose we choose $\{2,8\}$.

For MME states in $2\times 5$, \Eq{3} gives $r= 2$, so \Eq{58} and \Eq{59} should be enough to form an MME state [provided they satisfy \Eq{5}], which we now demonstrate.  First, list the chosen ME TGX tuples in coincidence form as
\begin{Equation}                      {60}
\begin{array}{*{20}l}
   {\{ 1,10\} } &\!\! { = \{ \{ 1,1\} ,\{ 2,5\} \} }  \\
   {\{ 2,8\} } &\!\! { = \{ \{ 1,2\} ,\{ 2,3\} \}. }  \\
\end{array}
\end{Equation}

Now, to get the $\delta_{k,l}$ in \Eq{5}, the indices of the modes of the bigger parts $B_m$ of \textit{each} extreme bipartition of these eigenstates need to correspond to \textit{different} levels in $B_m$ to get the spacewise orthogonality in $B_m$ which causes $\delta_{k,l}$. In $2\times 5$, there is only one unique extreme bipartition, so its unique $B_m$ is mode $2$. However, there are \textit{nominally two} extreme bipartitions which have \textit{mode-label lists} $B_1 =(2)$ and $B_2 =(2)$. Thus, listing all of the mode-$2$ indices of \Eq{60} taken together gives
\begin{Equation}                      {61}
\begin{array}{*{20}l}
   {B_1  = (2):} & {\{ \{ 1\} ,\{ 5\} |\{ 2\} ,\{ 3\} \} }  \\
   {B_2  = (2):} & {\{ \{ 1\} ,\{ 5\} |\{ 2\} ,\{ 3\} \}, }  \\
\end{array}
\end{Equation}
where bars separate sets from different ME TGX tuples. (This format helps relate it to the multipartite example.)

Since \Eq{51} needs to be satisfied for \textit{each} mode $m$, each line of \Eq{61} must contain different levels in each set, since only the mode-$B_m$ indices of \Eq{60} are involved in the \textit{partial trace} in \Eq{51} for mode $m$, while $\delta_{k,l}$ is satisfied by insisting that the levels separated by vertical bars in \Eq{61} are different. In fact, since each ME TGX tuple in \Eq{60} is made of spacewise orthogonal levels, we are guaranteed that the set of levels \textit{within} each bar-separated group in \Eq{61} will be different, \textit{so we can simply require that each line of \Eq{61} contains no repeated levels and ignore the bars}.  Since \Eq{61} passes this test, we could continue searching for new tuples; however we are actually finished, as we now explain.

The above explanation of how \Eq{61} relates to \Eq{51} shows why we cannot include more ME TGX tuples; the only unused mode-$2$ level in \Eq{61} is level $4$, but since all ME TGX tuples in $2\times 5$ have $L_* =2$ levels, there is no way to pick another ME TGX tuple that supplies a $4$ in mode $2$ without also incurring another level that is a repeat, since there are no remaining levels in mode $1$ beyond the already-used $1$ and $2$ and the new ME TGX tuple would need one of $\{1,2,3,5\}$ in mode $2$ in its other level because it represents an ME state and another $4$ would make it separable. Also, the maximum MME rank upper limit $\maxMMEranklim =2$ has already been reached, and that equivalently means there is no ``space'' left for another MME eigenstate.

(In general, if $\maxMMEranklim =1$, that is sufficient to conclude the system has no MME potential, but if $\maxMMEranklim \geq 2$, it does not necessarily imply MME potential (unless $N\!=\!2$); we have to search ME TGX tuples as we are doing.)

Thus, our constructed MME eigenstates in $2\times 5$ are found from \Eq{60} as
\vspace{-4pt}
\begin{Equation}                      {62}
\begin{array}{*{20}l}
   {\{ 1,10\} } &\!\! { \to |e_1 \rangle } &\!\! { \equiv \frac{1}{\sqrt{2}}(|1\rangle  + |10\rangle )}  \\
   {\{ 2,8\} } &\!\! { \to |e_2 \rangle } &\!\! { \equiv \frac{1}{\sqrt{2}}(|2\rangle  + |8\rangle ),}  \\
\end{array}
\end{Equation}
\vspace{-8pt}\\
which yield a family of TGX MME states as
\vspace{-6pt}
\begin{Equation}                      {63}
\rho _{\MME}  \equiv \lambda _1 |e_1 \rangle \langle e_1 | + \lambda _2 |e_2 \rangle \langle e_2 |;\;\,\lambda _k  \!\in\! (0,1),\;\, \sum\limits_{k = 1}^2 {\!\lambda _k }  \!=\! 1,
\end{Equation}
and by the MSD and \Eq{29}, this can be generalized to have \textit{non}TGX eigenstates (and thus generally be nonTGX itself) by any LU operation on the whole state as
\vspace{-4pt}
\begin{Equation}                      {64}
\rho _{\MME} ' \equiv (U^{(1)}  \otimes U^{(2)} )\rho _{\MME} (U^{(1)}  \otimes U^{(2)} )^\dag  ,
\end{Equation}
\vspace{-16pt}\\
where \smash{$U^{(m)}$} is any mode-$m$ unitary. Note that since LU operations are merely \textit{sufficient} to preserve entanglement of a given state, \Eq{64} is not the only way to generalize \Eq{63}; it would be more general to say
\vspace{-4pt}
\begin{Equation}                      {65}
\rho _{\MME} ' \equiv U_{\EPU} \rho _{\MME} U_{\EPU}^\dag  ,
\end{Equation}
\vspace{-16pt}\\
where \smash{$U_{\EPU}$} is any entanglement-preserving unitary (EPU), which allows nonlocal operations that preserve entanglement as well.

To verify that this method creates MME states, let $\rho  \equiv \rho _{\MME} '$ in \Eq{64} and relabel to get the form of \Eq{18} as
\vspace{-4pt}
\begin{Equation}                      {66}
\begin{array}{*{20}l}
   {\rho  = \sum\limits_{k = 1}^2 {\lambda_k |\Phi _k '\rangle \langle \Phi _k '|} ;} & {\left\{ {\begin{array}{*{20}l}
   {|\Phi _1 '\rangle } &\!\!\! { \equiv\! |e_1 \rangle  \!\equiv\! U_{\LU} |\Phi _{1,10}^ +  \rangle }  \\
   {|\Phi _2 '\rangle } &\!\!\! { \equiv\! |e_2 \rangle  \!\equiv\! U_{\LU} |\Phi _{2,8}^ +  \rangle, }  \\
\end{array}} \right.}  \\
\end{array}
\end{Equation}
where \smash{$U_{\LU}  \equiv U^{(1)}  \otimes U^{(2)} $} and\hsp{1} \smash{$|\Phi _{a,b}^ +  \rangle  \equiv \shiftmath{1.3pt}{\OneOverSqrt{2}}(|a\rangle  + |b\rangle )$}.  The reductions\hsp{-0.5} \smash{$\shiftmath{1.7pt}{\redx{\rho}{m}_{\hsp{-1}\shiftmath{-1pt}{|w_{j}\rangle}}}$}\hsp{-0.6} take\hsp{-0.6} the\hsp{-0.6} form\hsp{-0.6} of\hsp{-0.6} \Eq{22},\hsp{-1.5} for which we need \smash{$\tr_{\mbarsub}\hsp{-0.5} ( {|\Phi _{\shiftmath{1pt}{k}} '\rangle\hsp{-0.5} \langle \Phi _{\shiftmath{1pt}{l}} '|} )$}\hsp{3} [the work after \Eq{22} in \Sec{IV} derives the conditions for multipartite MME, and \textit{that} involves extreme bipartitions, but the mode-$m$ reductions are always given by the regular single-mode partial trace as in \Eq{22}]. Expanding the eigenstates in coincidence form,
\begin{Equation}                      {67}
\begin{array}{*{20}l}
   {|\Phi _1 '\rangle } &\!\! { = \frac{1}{{\sqrt 2 }}(|U_{:,1}^{(1)} \rangle |U_{:,1}^{(2)} \rangle  + |U_{:,2}^{(1)} \rangle |U_{:,5}^{(2)} \rangle )}  \\
   {|\Phi _2 '\rangle } &\!\! { = \frac{1}{{\sqrt 2 }}(|U_{:,1}^{(1)} \rangle |U_{:,2}^{(2)} \rangle  + |U_{:,2}^{(1)} \rangle |U_{:,3}^{(2)} \rangle ),}  \\
\end{array}
\end{Equation}
which shows the \textit{TGX indexing} of the multipartite Schmidt decomposition from \Eq{16}, here in a bipartite context.  Then the \smash{$\tr_{\mbarsub} ( {|\Phi _k '\rangle \langle \Phi _l '|} )$} for mode $1$ are
\begin{Equation}                      {68}
\begin{array}{*{20}l}
   {\tr_{\overline{1}} ( {|\Phi _1 '\rangle \langle \Phi _1 '|} )} &\!\! { = \frac{1}{2}( {|U_{:,1}^{(1)} \rangle \langle U_{:,1}^{(1)} | + |U_{:,2}^{(1)} \rangle \langle U_{:,2}^{(1)} |} )}  \\
   {\tr_{\overline{1}} ( {|\Phi _1 '\rangle \langle \Phi _2 '|} )} &\!\! { = \left( {\begin{array}{*{20}c}
   \cdot & \cdot  \\
   \cdot & \cdot  \\
\end{array}} \right)}  \\
   {\tr_{\overline{1}} \left( {|\Phi _2 '\rangle \langle \Phi _1 '|} \right)} &\!\! { = \left( {\begin{array}{*{20}c}
   \cdot & \cdot  \\
   \cdot & \cdot  \\
\end{array}} \right)}  \\
   {\tr_{\overline{1}} ( {|\Phi _2 '\rangle \langle \Phi _2 '|} )} &\!\! { = \frac{1}{2}( {|U_{:,1}^{(1)} \rangle \langle U_{:,1}^{(1)} | + |U_{:,2}^{(1)} \rangle \langle U_{:,2}^{(1)} |} ),}  \\
\end{array}
\end{Equation}
which simplifies to
\begin{Equation}                      {69}
\tr_{\overline{1}} ( {|\Phi _k '\rangle \langle \Phi _l '|} ) = \delta _{k,l} U^{(1)} \!\!\left( {\begin{array}{*{20}c}
   {\frac{1}{2}} & \cdot  \\
   \cdot & {\frac{1}{2}}  \\
\end{array}} \right)\! U^{(1)\dag }  = \delta _{k,l} {\textstyle \frac{1}{2}}I^{(1)} ,
\end{Equation}
where \smash{$\frac{1}{2}I^{(1)}$} is a state of minimum physical simultaneous reduction purity (MPSRP) of mode $1$ for $2\times 5$.

Now since the MME conditions only require \Eq{30} for the \textit{smaller} reduction, we do not actually need to achieve that for the bigger reduction. However, we \textit{do} want to see what those bigger reductions are, just to make sure everything works.  So for mode $2$, the \smash{$\tr_{\mbarsub} ( {|\Phi _k '\rangle \langle \Phi _l '|} )$} are
\begin{Equation}                      {70}
\begin{array}{*{20}l}
   {\tr_{\overline{2}} ( {|\Phi _1 '\rangle \langle \Phi _1 '|} )} &\!\! { = U^{(2)} \frac{1}{2}\!\left( {\scalemath{0.70}{\begin{array}{*{20}c}
   1 &  \cdot  &  \cdot  &  \cdot  &  \cdot   \\
    \cdot  &  \cdot  &  \cdot  &  \cdot  &  \cdot   \\
    \cdot  &  \cdot  &  \cdot  &  \cdot  &  \cdot   \\
    \cdot  &  \cdot  &  \cdot  &  \cdot  &  \cdot   \\
    \cdot  &  \cdot  &  \cdot  &  \cdot  & 1  \\
\end{array}}} \right)U^{(2)\dag } }  \\
   {\tr_{\overline{2}} ( {|\Phi _1 '\rangle \langle \Phi _2 '|} )} &\!\! { = U^{(2)} \frac{1}{2}\!\left( {\scalemath{0.70}{\begin{array}{*{20}c}
    \cdot  & 1 &  \cdot  &  \cdot  &  \cdot   \\
    \cdot  &  \cdot  &  \cdot  &  \cdot  &  \cdot   \\
    \cdot  &  \cdot  &  \cdot  &  \cdot  &  \cdot   \\
    \cdot  &  \cdot  &  \cdot  &  \cdot  &  \cdot   \\
    \cdot  &  \cdot  & 1 &  \cdot  &  \cdot   \\
\end{array}}} \right)U^{(2)\dag } }  \\
   {\tr_{\overline{2}} ( {|\Phi _2 '\rangle \langle \Phi _1 '|} )} &\!\! { = U^{(2)} \frac{1}{2}\!\left( {\scalemath{0.70}{\begin{array}{*{20}c}
    \cdot  &  \cdot  &  \cdot  &  \cdot  &  \cdot   \\
   1 &  \cdot  &  \cdot  &  \cdot  &  \cdot   \\
    \cdot  &  \cdot  &  \cdot  &  \cdot  & 1  \\
    \cdot  &  \cdot  &  \cdot  &  \cdot  &  \cdot   \\
    \cdot  &  \cdot  &  \cdot  &  \cdot  &  \cdot   \\
\end{array}}} \right)U^{(2)\dag } }  \\
   {\tr_{\overline{2}} ( {|\Phi _2 '\rangle \langle \Phi _2 '|} )} &\!\! { = U^{(2)} \frac{1}{2}\!\left( {\scalemath{0.70}{\begin{array}{*{20}c}
    \cdot  &  \cdot  &  \cdot  &  \cdot  &  \cdot   \\
    \cdot  & 1 &  \cdot  &  \cdot  &  \cdot   \\
    \cdot  &  \cdot  & 1 &  \cdot  &  \cdot   \\
    \cdot  &  \cdot  &  \cdot  &  \cdot  &  \cdot   \\
    \cdot  &  \cdot  &  \cdot  &  \cdot  &  \cdot   \\
\end{array}}} \right)U^{(2)\dag }\!. }  \\
\end{array}
\end{Equation}
Then, putting \Eq{69} into \Eq{22} reveals
\begin{Equation}                      {71}
\begin{array}{*{20}l}
   {\redx{\rho}{1}_{\!|w_{j}\rangle}} &\!\! {= \frac{{\sum\nolimits_{k = 1}^r {|U_{j,k} |^2 \lambda_k } \frac{1}{2}I^{(1)} }}{\rule{0pt}{8pt}{\sum\nolimits_{g = 1}^r {|U_{j,g} |^2 \lambda_g } }} = \frac{1}{2}I^{(1)} \,\,\forall j,\;\forall U,}  \\
\end{array}
\end{Equation}
while putting \Eq{70} into \Eq{22} reveals (after some work),
\begin{Equation}                      {72}
\begin{array}{*{20}l}
   {\redx{\rho}{2}_{\!|w_{j}\rangle}\!} &\!\!\! { =\! U^{(2)} \frac{1}{2}\!\!\left(\!\! {\rho _{\frac{\vphantom{|_{|_{-}}}{U_{\!j,\!1} \!\sqrt {\!\lambda_{\hsp{-0.5}1} } |1\rangle  \!+ U_{\!j,\!2} \!\sqrt {\!\lambda_2 } |2\rangle }}{\rule{0pt}{8pt}{\sqrt {\sum\nolimits_{k = 1}^r {|U_{\!j,\!k} |^2 \lambda_k } } }}}  \!\!+\!\! \rho _{\frac{\vphantom{|_{|_{-}}}{U_{\!j,\!2} \!\sqrt {\!\lambda_2 } |3\rangle  \!+\! U_{\!j,\!1} \!\sqrt {\!\lambda_{\hsp{-0.5}1} } |5\rangle }}{\rule{0pt}{8pt}{\sqrt {\sum\nolimits_{k = 1}^r {|U_{\!j,\!k} |^2 \lambda_k } } }}} }\!\! \right)\!\! U^{(2)\dag }\!, }  \\
\end{array}\!
\end{Equation}
where $\rho _{|\psi \rangle }  \equiv |\psi \rangle \langle \psi |$ and $r=2$. Note that \hyperlink{Mechanism:1}{Mechanism 1} acts in mode 1, while \hyperlink{Mechanism:2}{Mechanism 2} acts in mode 2.

The results of \Eq{71} and \Eq{72} are most illuminating. In \Eq{71}, we see that regardless of which decomposition unitary $U$ is used, all mode-$1$ reductions of all pure decomposition states are MPSRP states.  In \Eq{72}, the mode-$2$ reduction of each pure decomposition state of all decompositions is a rank-$2$ mixed state with equal eigenvalues (because the basis states in each of the pure decomposition states of \Eq{72} are spacewise orthogonal, guaranteeing that this is a spectral decomposition), and that this form holds for all decompositions is evidenced by the fact that the decomposition unitary $U$ plays the same role in the pure-state normalization for all states in these mixtures, while the two eigenvalues of $\frac{1}{2}$ are independent of $U$.

But here is the most important part: because the purity of \Eq{72} is an MPSRP of mode 2 in $2\times 5$ (which is independent of the local unitary $U^{(2)}$), that means that, together with \Eq{71}, \textit{both reductions of every pure decomposition state of every decomposition are as mixed as it is possible for them to be given a pure parent state, and that means that $\rho$ from \Eq{66} is MME}. 

\begin{figure}[H]
\centering
\includegraphics[width=1.00\linewidth]{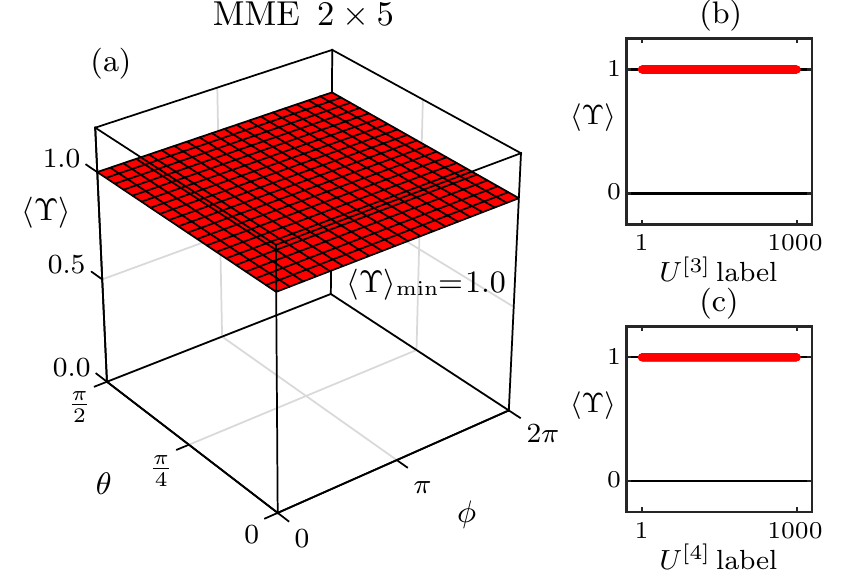}
\vspace{-16pt}
\caption{(color online) (a), (b), (c) show entanglement as minimum average ent \smash{$\langle\Upsilon\rangle_{\min}$} over many decompositions with decomposition unitaries \smash{$U^{[D]}\!\equiv\! U^{[D]}(\theta,\phi,\ldots)$} of sizes $D\!\in\!\{r,\ldots,r^2\}\!=\!\{2,3,4\}$ respectively for a rank-$2$ MME state \smash{$\rho_{\MME}$} in $2\!\times\! 5$ with arbitrary spectrum $\{\lambda_1,\lambda_2\}\!=\!\{0.7,0.3\}$ [\Sec{V.C} shows \smash{$U^{[2]}$} for (a)]. Each grid point or dot is $\langle\Upsilon\rangle$ for a different decomposition of $\rho_{\MME}$ [$400$ decompositions for (a), $1000$ decompositions for (b) and (c)]. This test strongly supports the claim that all pure decomposition states of all decompositions of $\rho_{\MME}$ have $\Upsilon\!=\!1$, since the minimum $\langle\Upsilon\rangle$ over all decompositions is \smash{$\langle\Upsilon\rangle_{\min}\!=\!1$}, so \smash{$\rho_{\MME}$} has the same entanglement as an ME pure state. These tests are not proofs; they are merely necessary tests of the proofs in \Sec{IV}.}
\label{fig:2}
\end{figure}

\Figure{2} demonstrates that $\rho_{\MME}$ of \Eq{63} [so also $\rho$ from \Eq{66} by LU equivalence] is MME by showing that its minimum average entanglement over many different decompositions is $1$.
\vspace{-7pt}

This example shows how to apply \Eq{5}, and illustrates how TGX indexing appears in the process. Now that we have shown how our method works for bipartite MME, we are ready for a multipartite example.
\subsection{\label{sec:V.B}Example for Multipartite MME}
In $2\times 2\times 2\times 2$, the extreme bipartitions sizes are $n_{S_m }  = 2$ and $n_{B_m }  = 8$ for $m=1,2,3,4$, so $\min\{\mathbf{n}_{B}\}=\min\{8,8,8,8\}=8$, and since $\mathbf{L}_*=\{2,4,6,8\}$ from \cite{HedE} or \Eq{6}, then $\min\{\mathbf{L}_{*}\}=2$, so by \Eq{5}, $\maxMMEranklim=\text{floor}(\frac{8}{2})=4>1$, so this system \textit{may} have MME potential if we can find at least two MME TGX tuples that satisfy \Eq{5}.

By exhaustive numerical search over all ME TGX tuples, using \Eq{5} and  $\mathcal{A}_{13}$ from \cite{HedE}, it was verified that $2\times 2\times 2\times 2$ can indeed host up to \textit{four} MME-compatible ME TGX tuples, so its maximal MME rank is $\maxMMErank=4$. However, to show that $2\times 2\times 2\times 2$ has MME potential, we merely need to find \textit{two} such tuples, so here we focus on the rank-$2$ case, which will also make this example simpler and easier to follow.

Therefore, (omitting the list of all ME TGX tuples for each starting level), if we choose $\{1,16\}$ as the first tuple, then searching the remaining tuple lists for starting levels $S_{L}\in 2,\ldots,15$ yields the following working set for the first two, with their coincidence form on the right as
\begin{Equation}                      {73}
\begin{array}{*{20}l}
   {\{ 1,16\} } &\!\! { = \{ \{ 1,1,1,1\} ,\{ 2,2,2,2\} \} }  \\
   {\{ 4,13\} } &\!\! { = \{ \{ 1,1,2,2\} ,\{ 2,2,1,1\} \} .}  \\
\end{array}
\end{Equation}

For now, suppose we do not know \Eq{73} is an MME-compatible pair.  To test its compatibility, we simply list out the ``big'' mode sets $B_m$ for $m=1,2,3,4$, and then extract only the $B_m$-indices from each tuple, again separating by bars to show their origin by tuple, which gives
\begin{Equation}                      {74}
\begin{array}{*{20}l}
   {B_1  = (2,3,4):} & {\{ \{ 1,1,1\} ,\{ 2,2,2\} | \{ 1,2,2\} ,\{ 2,1,1\} \} }  \\
   {B_2  = (1,3,4):} & {\{ \{ 1,1,1\} ,\{ 2,2,2\} | \{ 1,2,2\} ,\{ 2,1,1\} \} }  \\
   {B_3  = (1,2,4):} & {\{ \{ 1,1,1\} ,\{ 2,2,2\} | \{ 1,1,2\} ,\{ 2,2,1\} \} }  \\
   {B_4  = (1,2,3):} & {\{ \{ 1,1,1\} ,\{ 2,2,2\} | \{ 1,1,2\} ,\{ 2,2,1\} \}, }  \\
\end{array}
\end{Equation}
where, for example, each triplet in line 1 of \Eq{74} is taken from modes $2,3,4$ of each coincidence set of each ME TGX tuple in \Eq{73}. Here, it will help to use the indical register function \cite{HedE,HCor} to convert \Eq{74} to scalar indices, using $N'=3$, and letting $\mathbf{n}'$ be $(n_2 ,n_3 ,n_4 ),(n_1 ,n_3 ,n_4 ),(n_1 ,n_2 ,n_4 ),(n_1 ,n_2 ,n_3 )$ for each line of \Eq{74} respectively [all of which are $\mathbf{n}'=(2,2,2)$ for this example]. Thus \Eq{74} simplifies to
\begin{Equation}                      {75}
\begin{array}{*{20}l}
   {B_1  = (2,3,4):} & {\{ 1,8|4,5\} }  \\
   {B_2  = (1,3,4):} & {\{ 1,8|4,5\} }  \\
   {B_3  = (1,2,4):} & {\{ 1,8|2,7\} }  \\
   {B_4  = (1,2,3):} & {\{ 1,8|2,7\}, }  \\
\end{array}
\end{Equation}
which makes it easy to see that \textit{since each line of \Eq{75} contains no repeated levels, the ME TGX tuples in \Eq{73} are MME compatible}.  Again, this is our way of checking that this set of ME TGX tuples satisfies the constraint from \Eq{5} that $\tr[\tr_{B_m } (|\Phi _k \rangle \langle \Phi _l |)] = \delta _{k,l} \,\,\forall m \in 1, \ldots ,N$ for the case when $k\neq l$ which gives $\delta _{k,l}=0$ (the $\delta_{k,l}$ condition is spacewise orthonormality of all $B_m$ reductions of $\{|\Phi _k \rangle\}$). The case of $k=l$ for $\delta _{k,l}=1$ is guaranteed since each reduction of $|\Phi _k \rangle$ is already normalized.

From here, the construction of MME states from \Eq{73} is analogous to   \Eqs{62}{65}, so we are finished our simple rank-$2$ example of how to implement \Eq{5} for $2\times 2\times 2\times 2$.

If \Eq{73} had \textit{failed} this test (meaning at least one of the rows of \Eq{75} had at least one repeated index), then we would have to try another combination of ME TGX tuples.  At this point, since we have found two, we could continue to look for a third ME TGX tuple compatible with the first two.  In that case, each new compatibility test would \textit{start} with \Eq{75} and each row would have two \textit{more} levels appended to it, and then each row would need to have no repeats to verify a compatible third ME TGX tuple. However, this rank-$2$ example is sufficient to give a feeling for one way to implement an algorithm that realizes \Eq{5}.

The search over all ME TGX tuples is not always as simple as just weeding out ME TGX tuples that are not orthogonal to the first one and picking the next one in the list; doing so can lead to cases where the \textit{apparent} maximum MME rank is \textit{less than} the \textit{true} $\maxMMErank$. Therefore, the algorithm \textit{does} need to be an exhaustive search over all unique ME TGX tuples in general. However if $\maxMMEranklim$ is reached, that can allow the search to end early, unless the system's true $\maxMMErank$ is less than $\maxMMEranklim$.

To confirm that the eigenstates in \Eq{73} really \textit{do} support MME, again let an LU variation of that state be
\begin{Equation}                      {76}
\begin{array}{*{20}c}
   {\rho  = \sum\limits_{k = 1}^2 {\lambda_k |\Phi _k '\rangle \langle \Phi _k '|} ;} & {\left\{ {\begin{array}{*{20}l}
   {|\Phi _1 '\rangle } &\!\!\! { \equiv\! |e_1 \rangle  \!\equiv\! U_{\LU} |\Phi _{1,16}^ +  \rangle }  \\
   {|\Phi _2 '\rangle } &\!\!\! { \equiv\! |e_2 \rangle  \!\equiv\! U_{\LU} |\Phi _{4,13}^ +  \rangle, }  \\
\end{array}} \right.}  \\
\end{array}
\end{Equation}
where \smash{$|\Phi _{\shiftmath{0pt}{a,\!b}}^ +  \rangle  \equiv\! \shiftmath{1pt}{\OneOverSqrt{2}}(|a\rangle  + |b\rangle )$} and \smash{$U_{\LU}  \equiv U^{(1)}  \otimes U^{(2)}  \otimes$} \smash{$ U^{(3)}\!  \otimes\! U^{(4)} $}.\hsp{3} The eigenstates in coincidence form are
\begin{Equation}                      {77}
\scalemath{0.90}{\begin{array}{*{20}l}
   {|\Phi _1 '\rangle } &\!\! { =\! {\textstyle \frac{1}{{\sqrt 2 }}}(|U_{:,1}^{(1)} \rangle |U_{:,1}^{(2)} \rangle |U_{:,1}^{(3)} \rangle |U_{:,1}^{(4)} \rangle  \!+\! |U_{:,2}^{(1)} \rangle |U_{:,2}^{(2)} \rangle |U_{:,2}^{(3)} \rangle |U_{:,2}^{(4)} \rangle )}  \\
   {|\Phi _2 '\rangle } &\!\! { =\! {\textstyle \frac{1}{{\sqrt 2 }}}(|U_{:,1}^{(1)} \rangle |U_{:,1}^{(2)} \rangle |U_{:,2}^{(3)} \rangle |U_{:,2}^{(4)} \rangle  \!+\! |U_{:,2}^{(1)} \rangle |U_{:,2}^{(2)} \rangle |U_{:,1}^{(3)} \rangle |U_{:,1}^{(4)} \rangle ),}  \\
\end{array}}
\end{Equation}
and in \textit{this} system, $S_m =m\;\forall m$ and \textit{each} $\tr_{\overline{m}} (|\Phi _k '\rangle \langle \Phi _l '|)$ has the same output as \Eq{68} for all modes $m$, yielding
\begin{Equation}                      {78}
\begin{array}{*{20}l}
   {\tr_{\overline{1}} (|\Phi _k '\rangle \langle \Phi _l '|)} &\!\! { = \delta _{k,l} U^{(1)} \left( {\begin{array}{*{20}c}
   {\frac{1}{2}} & \cdot  \\
   \cdot & {\frac{1}{2}}  \\
\end{array}} \right)U^{(1)\dag } } &\!\! { = \delta _{k,l} \frac{1}{2}I^{(1)} }  \\
   {\tr_{\overline{2}} (|\Phi _k '\rangle \langle \Phi _l '|)} &\!\! { = \delta _{k,l} U^{(2)} \left( {\begin{array}{*{20}c}
   {\frac{1}{2}} & \cdot  \\
   \cdot & {\frac{1}{2}}  \\
\end{array}} \right)U^{(2)\dag } } &\!\! { = \delta _{k,l} \frac{1}{2}I^{(2)} }  \\
   {\tr_{\overline{3}} (|\Phi _k '\rangle \langle \Phi _l '|)} &\!\! { = \delta _{k,l} U^{(3)} \left( {\begin{array}{*{20}c}
   {\frac{1}{2}} & \cdot  \\
   \cdot & {\frac{1}{2}}  \\
\end{array}} \right)U^{(3)\dag } } &\!\! { = \delta _{k,l} \frac{1}{2}I^{(3)} }  \\
   {\tr_{\overline{4}} (|\Phi _k '\rangle \langle \Phi _l '|)} &\!\! { = \delta _{k,l} U^{(4)} \left( {\begin{array}{*{20}c}
   {\frac{1}{2}} & \cdot  \\
   \cdot & {\frac{1}{2}}  \\
\end{array}} \right)U^{(4)\dag } } &\!\! { = \delta _{k,l} \frac{1}{2}I^{(4)}, }  \\
\end{array}
\end{Equation}
so in $2\times 2\times 2\times 2$, the states on the right in \Eq{78} next to $\delta_{k,l}$ are all identical states of MPSRP for all $k,l\in 1,2$. This is different from the $2\times 5$ example where we had a single largest mode.  \{In general, multipartite states with a \textit{single largest mode} will have MPSRPs which prevent some reductions from reaching ideal maximal mixing given a pure ME parent state, as stated in \Eq{40}, based on \cite{HedE}.\}  Then, putting \Eq{78} into \Eq{22} yields\pagebreak 
\\\vspace{-18pt}
\begin{Equation}                      {79}
\begin{array}{*{20}l}
   {\redx{\rho}{1}_{|w_{j}\rangle} } &\!\! {= \frac{{\sum\nolimits_{k = 1}^r {|U_{j,k} |^2 \lambda_k } \frac{1}{2}I^{(1)} }}{\rule{0pt}{8pt}{\sum\nolimits_{g = 1}^r {|U_{j,g} |^2 \lambda_g } }} } &\!\! { = \frac{1}{2}I^{(1)} \,\,\forall j,\;\forall U}  \\
   {\redx{\rho}{2}_{|w_{j}\rangle} } &\!\! {= \frac{{\sum\nolimits_{k = 1}^r {|U_{j,k} |^2 \lambda_k } \frac{1}{2}I^{(2)} }}{\rule{0pt}{8pt}{\sum\nolimits_{g = 1}^r {|U_{j,g} |^2 \lambda_g } }} } &\!\! { = \frac{1}{2}I^{(2)} \,\,\forall j,\;\forall U\rule{0pt}{15pt}}  \\
   {\redx{\rho}{3}_{|w_{j}\rangle} } &\!\! {= \frac{{\sum\nolimits_{k = 1}^r {|U_{j,k} |^2 \lambda_k } \frac{1}{2}I^{(3)} }}{\rule{0pt}{8pt}{\sum\nolimits_{g = 1}^r {|U_{j,g} |^2 \lambda_g } }} } &\!\! { = \frac{1}{2}I^{(3)} \,\,\forall j,\;\forall U\rule{0pt}{15pt}}  \\
   {\redx{\rho}{4}_{|w_{j}\rangle} } &\!\! {= \frac{{\sum\nolimits_{k = 1}^r {|U_{j,k} |^2 \lambda_k } \frac{1}{2}I^{(4)} }}{\rule{0pt}{8pt}{\sum\nolimits_{g = 1}^r {|U_{j,g} |^2 \lambda_g } }} } &\!\! { = \frac{1}{2}I^{(4)} \,\,\forall j,\;\forall U,\rule{0pt}{15pt}}  \\
\end{array}
\end{Equation}
where $r=2$, which shows that $U$ cancels analogously to \Eq{71} and means that every pure decomposition state $|w_{j}\rangle$ of all decompositions has all of its modes in states of MPSRP, so all decompositions consist of ME pure states, and therefore \Eq{76} is MME in $2\times 2\times 2\times 2$.

\Figure{1} (shown in the summary of results in \Sec{II}) shows that the  $2\times 2\times 2\times 2$ state $\rho$ of \Eq{76} (omitting $U_{\LU}$ since it cannot affect entanglement, and using an arbitrary rank-$2$ spectrum) is MME by testing its minimum average entanglement over a large number of different decompositions, and shows the same tests for a nonMME entangled state and a separable state in the same system for comparison.  It also shows the same tests using more decomposition pure states than the rank, by making $1000$ arbitrary decomposition unitaries $U$ of dimensions $3$ and $4$ to show that even using up to $r^2$ decomposition states, we cannot find any decompositions with a lower minimum average entanglement than $1$.  Thus, this test strongly supports our claim that \Eq{76} is a multipartite MME state.

\begin{figure}[H]
\centering
\includegraphics[width=1.00\linewidth]{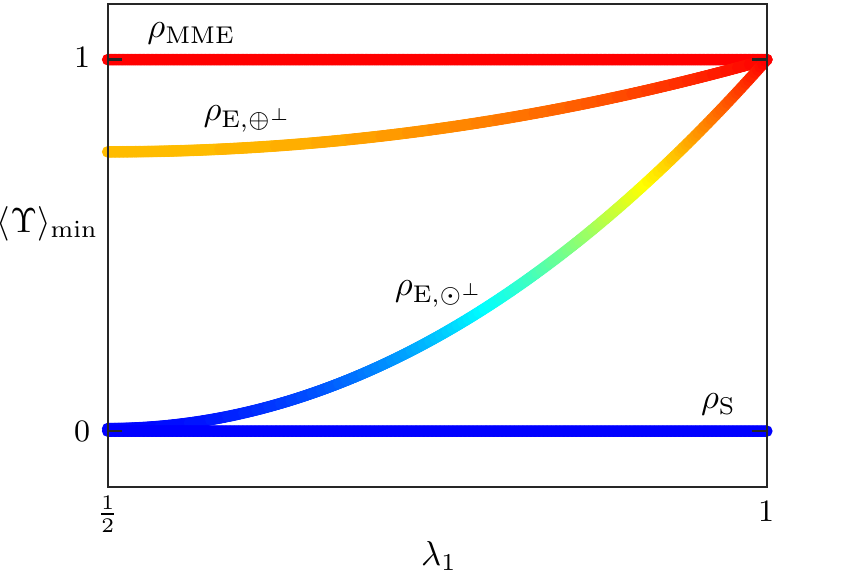}
\vspace{-13pt}
\caption{(color online) Plot of entanglement versus largest eigenvalue $\lambda_1$, where entanglement is given by minimum average ent \smash{$\langle\Upsilon\rangle_{\min}$} over many decompositions with a grid-search over all $2$-level decomposition unitaries \smash{$U^{[2]}(\theta,\phi)$} (see \Sec{V.C} for \smash{$U^{[2]}$}) at a resolution of $20$ points per angle, so each point here is \smash{$\langle\Upsilon\rangle_{\min}$} over $400$ decompositions, for the rank-$2$ states in $2\times 2\times 2\times 2$ from \Eq{80}.  State families \smash{$\rho_{\MME}$}, \smash{$\rho_{\text{E},\oplus^{\perp}}$}, \smash{$\rho_{\text{E},\odot^{\perp}}$}, \smash{$\rho_{\text{S}}$} were each separately confirmed to produce the curves as they appear and are labeled, so they are shown here with the same color map as \Fig{1} and \Fig{2} for continuity. Each state's curve shows its entanglement for $500$ different spectra. This illustrates that MME states have entanglement $1$ for all spectra in an MME-compatible rank. It also shows that not all spacewise orthogonal ME TGX tuples can form an MME-compatible set; they must satisfy the constraints of \Eq{5}, and when they do not, the entanglement will be less than $1$, as seen for \smash{$\rho_{\text{E},\oplus^{\perp}}$}.  Meanwhile, \smash{$\rho_{\text{E},\odot^{\perp}}$} shows the need for spacewise orthogonality since it uses self-space orthogonality and fails to be MME as well.}
\label{fig:3}
\end{figure}

To show the independence of MME states from spectrum (other than by rank), \Figure{3} shows entanglement as minimum average ent $\langle\Upsilon\rangle_{\min}$ over all decompositions versus the largest eigenvalue $\lambda_1$ for the rank-$2$ MME state family $\rho$ we have just constructed in \Eq{76} (including the pure-state limit here for context, and again ignoring $U_{\LU}$ since it does not affect entanglement) and several other states for comparison.  The states we compare here are
\begin{Equation}                      {80}
\begin{array}{*{20}l}
   {\rho _{\MME} } &\!\! { \equiv \lambda _1 \rho _{|\Phi _{1,16}^ +  \rangle }  + \lambda _2 \rho _{|\Phi _{4,13}^ +  \rangle } }  \\
   {\rho _{\text{E}, {\oplus^\perp} } } &\!\! { \equiv \lambda _1 \rho _{|\Phi _{1,16}^ +  \rangle }  + \lambda _2 \rho _{|\Phi _{2,15}^ +  \rangle } }  \\
   {\rho _{\text{E},{\odot^\perp}  } } &\!\! { \equiv \lambda _1 \rho _{|\Phi _{1,16}^ +  \rangle }  + \lambda _2 \rho _{|\Phi _{1,16}^ -  \rangle } }  \\
   {\rho _{\text{S}} } &\!\! { \equiv \lambda _1 \rho _{|1\rangle }  + \lambda _2 \rho _{|16\rangle } ,}  \\
\end{array}
\end{Equation}
where \smash{$\rho _{|\psi \rangle }  \equiv |\psi \rangle \langle \psi |$} and \smash{$|\Phi _{a,b}^ \pm  \rangle  \equiv \shiftmath{1pt}{\OneOverSqrt{2}}(|a\rangle  \pm |b\rangle )$}. Here, $\rho _{\text{E}, {\oplus^\perp} }$\hsp{-0.5} is\hsp{-0.5} an\hsp{-0.5} entangled\hsp{-0.5} state\hsp{-0.5} family\hsp{-0.5} whose ME TGX eigenstates are spacewise orthogonal (since these ME TGX tuples do not reside in the same subspace). Since the entanglement of $\rho _{\text{E}, {\oplus^\perp} }$ is generally less than $1$, this shows that we cannot simply choose \textit{any} spacewise orthogonal set of ME TGX tuples to get an MME state; the tuples have to also obey \Eq{5}.  Meanwhile, $\rho _{\text{E},{\odot^\perp}  }$ is an entangled state family whose eigenstates are \textit{self-space orthogonal}, meaning they share the same subspace and their orthogonality comes from cancellation in their dot product. Since the entanglement of $\rho _{\text{E},{\odot^\perp}  }$ is also generally less than $1$, this shows that we truly need multiple spacewise orthogonal ME TGX states to achieve MME.  Finally, $\rho _{\text{S}}$ is a separable state family shown for reference.  Furthermore, \Fig{3} shows that since MME states are special kinds of MEMS, they have unit-entanglement plateaus in MEMS curves of entanglement versus any function of spectrum.
\vspace{-16pt}
\subsection{\label{sec:V.C}Tables of $\maxMMErank$ and MME Eigenstates}
Here we present the tabulated results of applying \Eq{5} to many systems. Again, we include bipartite systems for completeness, and in some cases, we provide additional tables to show more multipartite results.

First, \Table{1} shows the maximal MME rank $\maxMMErank$ to indicate the MME potential for the first $28$ dimensions, to parallel the tables in \cite{HedE}. This table also shows that not all systems permit MME.

Then, \Table{2} shows examples of eigenstate sets for TGX MME states for all MME-compatible ranks in systems with MME potential in the first $28$ dimensions. These can all be generalized to \textit{non}TGX states by LU operations, or even by a more general EPU. (Note that the eigenstates of nonTGX MME states are \textit{not} generally \textit{spacewise} orthogonal; they are orthogonal in more general ways. The spacewise orthogonality requirement in \Eq{5} only applies to the TGX core states of the multipartite Schmidt decomposition of the eigenstates).
\begin{table}[H]
\caption{\label{tab:1}MME potential in the first $28$ dimensions of multipartite systems, indicated by \textit{maximal MME rank} $\maxMMErank$ defined in \Eq{5}. $\maxMMErank=1$ means that only \textit{pure} states can be ME (no MME potential), whereas $\maxMMErank\geq 2$ means MME is possible up to rank $\maxMMErank$ (so there is MME potential).}
\begin{ruledtabular}
\begin{tabular}{|c|c|c|c|@{${\kern -4pt}$}c|c|c|}
$n$ & $n_{1}\times\cdots\times n_{N}$ & $\maxMMErank$ & &{\kern 2pt}$n$ & $n_{1}\times\cdots\times n_{N}$ & $\maxMMErank$ \\[0.5mm]
\cline{1-3} \cline{5-7}
$4\rule{0pt}{10pt}$ & $2\times 2$ & $1$ & &{\kern 2pt}$20$ & $2\times 10$ & $5$\\[0.5mm]
\cline{1-3} \cline{5-7}
$6\rule{0pt}{10pt}$ & $2\times 3$ & $1$ & &{\kern 2pt}$20$ & $4\times 5$ & $1$\\[0.5mm]
\cline{1-3} \cline{5-7}
$8\rule{0pt}{10pt}$ & $2\times 4$ & $2$ & &{\kern 2pt}$20$ & $2\times 2\times 5$ & $1$\\[0.5mm]
\cline{1-3} \cline{5-7}
$8\rule{0pt}{10pt}$ & $2\times 2\times 2$ &$1$ & &{\kern 2pt}$21$ & $3\times 7$ & $2$\\[0.5mm]
\cline{1-3} \cline{5-7}
$9\rule{0pt}{10pt}$ & $3\times 3$ & $1$ & &{\kern 2pt}$22$ & $2\times 11$ & $5$\\[0.5mm]
\cline{1-3} \cline{5-7}
$10\rule{0pt}{10pt}$ & $2\times 5$ & $2$ & &{\kern 2pt}$24$ & $2\times 12$ & $6$\\[0.5mm]
\cline{1-3} \cline{5-7}
$12\rule{0pt}{10pt}$ & $2\times 6$ & $3$ & &{\kern 2pt}$24$ & $3\times 8$ & $2$\\[0.5mm]
\cline{1-3} \cline{5-7}
$12\rule{0pt}{10pt}$ & $3\times 4$ & $1$ & &{\kern 2pt}$24$ & $4\times 6$ & $1$\\[0.5mm]
\cline{1-3} \cline{5-7}
$12\rule{0pt}{10pt}$ & $2\times 2\times 3$ & $1$ & &{\kern 2pt}$24$ & $2\times 2\times 6$ & $1$\\[0.5mm]
\cline{1-3} \cline{5-7}
$14\rule{0pt}{10pt}$ & $2\times 7$ & $3$ & &{\kern 2pt}$24$ & $2\times 3\times 4$ & $1$\\[0.5mm]
\cline{1-3} \cline{5-7}
$15\rule{0pt}{10pt}$ & $3\times 5$ & $1$ & &{\kern 2pt}$24$ & $2\times 2\times 2\times 3$ & $1$\\[0.5mm]
\cline{1-3} \cline{5-7}
$16\rule{0pt}{10pt}$ & $2\times 8$ & $4$ & &{\kern 2pt}$25$ & $5\times 5$ & $1$\\[0.5mm]
\cline{1-3} \cline{5-7}
$16\rule{0pt}{10pt}$ & $4\times 4$ & $1$ & &{\kern 2pt}$26$ & $2\times 13$ & $6$\\[0.5mm]
\cline{1-3} \cline{5-7}
$16\rule{0pt}{10pt}$ & $2\times 2\times 4$ & $1$ & &{\kern 2pt}$27$ & $3\times 9$ & $3$\\[0.5mm]
\cline{1-3} \cline{5-7}
$16\rule{0pt}{10pt}$ & $2\times 2\times 2\times 2$ & $4$ & &{\kern 2pt}$27$ & $3\times 3\times 3$ & $3$\\[0.5mm]
\cline{1-3} \cline{5-7}
$18\rule{0pt}{10pt}$ & $2\times 9$ & $4$ & &{\kern 2pt}$28$ & $2\times 14$ & $7$\\[0.5mm]
\cline{1-3} \cline{5-7}
$18\rule{0pt}{10pt}$ & $3\times 6$ & $2$ & &{\kern 2pt}$28$ & $4\times 7$ & $1$\\[0.5mm]
\cline{1-3} \cline{5-7}
$18\rule{0pt}{10pt}$ & $2\times 3\times 3$ & $1$ & &{\kern 2pt}$28$ & $2\times 2\times 7$ & $1$\\
\end{tabular}
\end{ruledtabular}
\end{table}
From \Table{1} and \Table{2}, we see that the first $n=28$ dimensions only include \textit{two} systems of distinctly multipartite ($N\geq 3$) MME; $2\times 2\times 2\times 2$ and $3\times 3\times 3$.

Therefore, \Table{3} extends our study to the first $n=36$ dimensions, and focuses only on systems of three or more modes ($N\geq 3$) (omitting two-mode results, and repeating the first two distinctly multipartite systems with MME potential from \Table{1} for completeness).

Then, \Table{4} focuses only on the systems from \Table{3} that can have MME, showing examples of MME-compatible eigenstates represented as ME TGX tuples.

Finally, \Table{5} shows $\maxMMEranklim$, $\maxMMErank$, and examples of MME-compatible eigenstates for $N$-qubit systems, where we represent the systems as $2^{\otimes N}\equiv 2_{1}\times\cdots\times 2_{N}$, so for example, $2^{\otimes 2}\equiv 2\times 2$, and $2^{\otimes 3}\equiv 2\times 2\times2 $, etc.

Additionally, for \textit{all} of the MME-supporting systems in \textit{all} of these tables, we did numerical tests to verify the MME effect.  This was done in two ways. First, for a minimal test, two of the example eigenstates and an arbitrary pair of nonzero eigenvalues were used to generate an MME state, which was then minimally decomposed using a rank-$2$ decomposition unitary $U\equiv U^{[2]}$ and searched over the two degrees of freedom that appear in $U_{j,k}U_{j,l}^{*}$ in \Eq{22}. Thus if $U\equiv(\hphantom{|}_{-b^*}^{\hphantom{-}a}\,\hphantom{|}_{a^*}^{b})$ with $(a,b)\equiv(\cos(\theta),\sin(\theta)e^{i\chi})$ where $\theta\in[0,\frac{\pi}{2}]$ and $\chi\in[0,2\pi)$, then this allows us to search $\theta$ and $\chi$ as a necessary test to see what the minimum average entanglement is.
\begin{table}[H]
\caption{\label{tab:2}Examples of MME states in all MME-hosting systems up to $n=28$ up to maximal MME rank $\maxMMErank$, where $\rho_{\MME}$ is an MME TGX state with eigenstates shown represented as ME TGX tuples in scalar-index form.  For example, in $2\times 6$, the rank-$3$ MME represented is $\rho_{\MME}=\sum\nolimits_{j=1}^{3}\lambda_{j}|e_{j}\rangle\langle e_{j}|$ with ME eigenstates \smash{$|e_{1}\rangle=\frac{1}{\sqrt{2}}(|1\rangle+|12\rangle)$}, \smash{$|e_{2}\rangle=\frac{1}{\sqrt{2}}(|2\rangle+|9\rangle)$}, and \smash{$|e_{3}\rangle=\frac{1}{\sqrt{2}}(|4\rangle+|11\rangle)$}, while rank-$2$ examples can be made from any pair of these eigenstates such as $\rho_{\MME}=\lambda_{1}|e_{1}\rangle\langle e_{1}|+\lambda_{2}|e_{2}\rangle\langle e_{2}|$. Such states are MME for all spectra and EPU variations.}
\vspace{-4pt}
\begin{ruledtabular}
\begin{tabular}{|c|c|c|}
$n$ & $n_{1}\!\times\!\cdots\!\times\! n_{N}$& Levels of Eigenstates of $\rho_{\MME}$\\[0.5mm]
\hline 
$8\rule{0pt}{10pt}$ & $2\times 4$ & $\{1,8\}, \{2,7\}\,\,$ \\[0.5mm]
\hline
$10\rule{0pt}{10pt}$ & $2\times 5$ & $\{1,10\}, \{2,8\}\,\,$ \\[0.5mm]
\hline
$12\rule{0pt}{10pt}$ & $2\times 6$ & $\{1,12\}, \{2,9\}, \{4,11\}\,\,$ \\[0.5mm]
\hline
$14\rule{0pt}{10pt}$ & $2\times 7$ & $\{1,14\}, \{2,10\}, \{4,12\}\,\,$ \\[0.5mm]
\hline
$16\rule{0pt}{10pt}$ & $2\times 8$ & $\{1,16\}, \{2,11\}, \{4,13\}, \{6,15\}\,\,$ \\[0.5mm]
\hline
$16\rule{0pt}{10pt}$ & $2\times 2\times 2\times 2$ & $\{1,16\}, \{4,13\}, \{6,11\}, \{7,10\}\,\,$ \\[0.5mm]
\hline
$18\rule{0pt}{10pt}$ & $2\times 9$ & $\{1,18\}, \{2,12\}, \{4,14\}, \{6,16\}\,\,$ \\[0.5mm]
\hline 
$18\rule{0pt}{10pt}$ & $3\times 6$ & $\{1,8,18\}, \{3,10,17\}\,\,$ \\[0.5mm]
\hline
$20\rule{0pt}{10pt}$ & $2\times 10$ & $\{1,20\}, \{2,13\}, \{4,15\}, \{6,17\},\{8,19\}\,\,$ \\[0.5mm]
\hline 
$21\rule{0pt}{10pt}$ & $3\times 7$ & $\{1,9,21\}, \{3,11,19\}\,\,$ \\[0.5mm]
\hline
$22\rule{0pt}{10pt}$ & $2\times 11$ & $\{1,22\}, \{2,14\}, \{4,16\}, \{6,18\},\{8,20\}\,\,$ \\[0.5mm]
\hline
$\begin{array}{*{20}c}
   {24\rule{0pt}{10pt}}  \\
   {}  \\
\end{array}$ & $\begin{array}{*{20}c}
   {2\times 12 \rule{0pt}{10pt}}  \\
   {}  \\
\end{array}$ & $\begin{array}{*{20}l}
   {\{1,24\}, \{2,15\}, \{4,17\}, \{6,19\}, \{8,21\},}  \\
   {\{10,23\} }  \\
\end{array}$ \\[0.5mm]
\hline 
$24\rule{0pt}{10pt}$ & $3\times 8$ & $\{1,10,24\}, \{3,12,21\}\,\,$ \\[0.5mm]
\hline
$\begin{array}{*{20}c}
   {26\rule{0pt}{10pt}}  \\
   {}  \\
\end{array}$ & $\begin{array}{*{20}c}
   {2\times 13 \rule{0pt}{10pt}}  \\
   {}  \\
\end{array}$ & $\begin{array}{*{20}l}
   {\{1,26\}, \{2,16\}, \{4,18\}, \{6,20\}, \{8,22\},}  \\
   {\{10,24\} }  \\
\end{array}$ \\[0.5mm]
\hline 
$27\rule{0pt}{10pt}$ & $3\times 9$ & $\{1,11,27\}, \{3,13,23\}, \{6,16,26\}\,\,$ \\[0.5mm]
\hline 
$27\rule{0pt}{10pt}$ & $3\times 3\times 3$ & $\{1,14,27\}, \{6,16,20\}, \{8,12,22\}\,\,$ \\[0.5mm]
\hline
$\begin{array}{*{20}c}
   {28\rule{0pt}{10pt}}  \\
   {}  \\
\end{array}$ & $\begin{array}{*{20}c}
   {2\times 14 \rule{0pt}{10pt}}  \\
   {}  \\
\end{array}$ & $\begin{array}{*{20}l}
   {\{1,28\}, \{2,17\}, \{4,19\}, \{6,21\}, \{8,23\},}  \\
   {\{10,25\}, \{12,27\} }  \\
\end{array}$ \\
\end{tabular}
\end{ruledtabular}
\end{table}
\phantom{~}\vspace{-18pt}

It should be emphasized that such a search of a finite number of decompositions is not a proof, but rather it is a necessary test to check that there is no apparent failure of the proof of MME which is in \Sec{IV}.  

\Figure{1} shows the results of this minimal test for an MME state in $2\times 2\times 2\times 2$ with the eigenstates of \Eq{73}.  The fact that we were not able to find a single decomposition that had an average entanglement less than $1$ while doing a search over a uniform distribution of the degrees of freedom of $U$ is very persuasive evidence that the proof in \Sec{IV} is not wrong.

For more extensive numerical tests, for each system in \Tables{1}{5}, we generated $100$ arbitrary decomposition unitaries $U^{[D]}$ with dimensions $D\in r,\ldots,r^2$ for \textit{each} MME-compatible rank $r\in 2,\ldots,\maxMMErank$ and confirmed the MME behavior.  Subfigures \Fig[b]{1} and \Fig[c]{1} show an example of these tests for a rank-$2$ MME state in $2\times 2\times 2\times 2$ (with the $D\geq3$ tests over $1000$ decompositions there), and show that we cannot find a decomposition with lower average entanglement than $1$, so this also \makebox[\linewidth][s]{supports the proof in \Sec{IV}.  These same tests were}
\begin{table}[H]
\caption{\label{tab:3}MME potential in the first $36$ dimensions of all multipartite systems of at least 3 modes ($N\geq 3$), indicated by maximal MME rank $\maxMMErank$ from \Eq{5} so that $\maxMMErank=1$ means only \textit{pure} states can be ME, whereas $\maxMMErank\geq 2$ means MME is possible up to rank $\maxMMErank$. Values of $\min\{\mathbf{L}_{*}\}$ and the loose maximal-MME limit $\maxMMEranklim$ are shown for reference. (For $n\leq 28$, only systems with MME are shown to save space.)}
\vspace{-4pt}
\begin{ruledtabular}
\begin{tabular}{|c|c|c|c|c|}
{ $n$ } & { $n_{1}\times\cdots\times n_{N}$ } & { $\min\{\mathbf{L}_{*}\}$ } &  { $\maxMMEranklim$ } & { $\maxMMErank$ } \\[0.5mm]%
\hline
{ $16$\rule{0pt}{10pt} } & { $2\times 2\times 2\times 2$ } & { $2$ } & { $4$ } & { $4$ } \\[0.5mm]%
\hline
{ $27$\rule{0pt}{10pt} } & { $3\times 3\times 3$ } & { $3$ } & { $3$ } & { $3$ } \\[0.5mm]%
\hline
{ $30$\rule{0pt}{10pt} } & { $2\times 3\times 5$ } & { $6$ } & { $1$ } & { $1$ } \\[0.5mm]%
\hline
{ $32$\rule{0pt}{10pt} } & { $2\times 2\times 8$ } & { $4$ } & { $2$ } & { $2$ } \\[0.5mm]%
\hline
{ $32$\rule{0pt}{10pt} } & { $2\times 4\times 4$ } & { $4$ } & { $2$ } & { $2$ } \\[0.5mm]%
\hline
{ $32$\rule{0pt}{10pt} } & { $2\times 2\times 2\times 4$ } & { $4$ } & { $2$ } & { $2$ } \\[0.5mm]%
\hline
{ $32$\rule{0pt}{10pt} } & { $2\times 2\times 2\times 2\times 2$ } & { $2$ } & { $8$ } & { $5$ } \\[0.5mm]%
\hline
{ $36$\rule{0pt}{10pt} } & { $2\times 2\times 9$ } & { $4$ } & { $2$ } & { $2$ } \\[0.5mm]%
\hline
{ $36$\rule{0pt}{10pt} } & { $2\times 3\times 6$ } & { $6$ } & { $1$ } & { $1$ } \\[0.5mm]%
\hline
{ $36$\rule{0pt}{10pt} } & { $3\times 3\times 4$ } & { $9$ } & { $1$ } & { $1$ } \\[0.5mm]%
\hline
{ $36$\rule{0pt}{10pt} } & { $2\times 2\times 3\times 3$ } & { $6$ } & { $2$ } & { $2$ } \\
\end{tabular}
\end{ruledtabular}
\end{table}
\vspace{-14pt}
\begin{table}[H]
\caption{\label{tab:4}Examples of MME states in all MME-hosting systems up to $n=36$ of at least $3$ modes ($N\geq 3$) up to maximal MME rank $\maxMMErank$, where $\rho_{\MME}$ is an MME TGX state with eigenstates shown represented as ME TGX tuples in scalar-index form, as explained in \Table{2}. Such states are MME for all spectra and EPU variations.}
\vspace{-4pt}
\begin{ruledtabular}
\begin{tabular}{|c|c|c|}
{ $n$ } & { $n_{1}\!\times\!\cdots\!\times\! n_{N}$ } & { Levels of Eigenstates of $\rho_{\MME}$ }\\[0.5mm]
\hline
{ $16$\rule{0pt}{10pt} } & { $2\times 2\times 2\times 2$ } & { $\{1,16\}, \{4,13\}, \{6,11\}, \{7,10\}\,\,$ } \\[0.5mm]%
\hline
{ $27$\rule{0pt}{10pt} } & { $3\times 3\times 3$ } & { $\{1,14,27\}, \{6,16,20\}, \{8,12,22\}\,\,$ } \\[0.5mm]%
\hline
{ $32$\rule{0pt}{10pt} } & { $2\times 2\times 8$ } & { $\{1,10,19,32\},\{4,13,22,31\}$ } \\[0.5mm]%
\hline
{ $32$\rule{0pt}{10pt} } & { $2\times 4\times 4$ } & { $\{1,6,27,32\},\{12,15,18,21\}$ } \\[0.5mm]%
\hline
{ $32$\rule{0pt}{10pt} } & { $2\times 2\times 2\times 4$ } & { $\{1,6,27,32\},\{10,13,20,23\}$ } \\[0.5mm]%
\hline
$\begin{array}{*{20}c}
   {32\rule{0pt}{10pt}}  \\
   {}  \\
\end{array}$ & $\begin{array}{*{20}c}
   {2\times 2\times 2\times 2\times 2 \rule{0pt}{10pt}}  \\
   {}  \\
\end{array}$ & $\begin{array}{*{20}l}
   { \{1,32\},\{4,29\},\{6,27\},\{10,23\}, }  \\
   { \{15,18\} }  \\
\end{array}$ \\[0.5mm]%
\hline
{ $36$\rule{0pt}{10pt} } & { $2\times 2\times 9$ } & { $\{1,11,21,36\},\{4,14,24,34\}$ } \\[0.5mm]%
\hline
$\begin{array}{*{20}c}
   {36\rule{0pt}{10pt}}  \\
   {}  \\
\end{array}$ & $\begin{array}{*{20}c}
   {2\times 2\times 3\times 3 \rule{0pt}{10pt}}  \\
   {}  \\
\end{array}$ & $\begin{array}{*{20}c}
   { \{1,5,9,28,32,36\}, }  \\
   { \{11,15,16,20,24,25\} }  \\
\end{array}$ \\
\end{tabular}
\end{ruledtabular}
\end{table}
\vspace{-6pt}
{\noindent}also applied in \Table{5} to systems $2^{\otimes 2}$, $2^{\otimes 3}$, $2^{\otimes 4}$, $2^{\otimes 5}$. For systems $2^{\otimes 6}$, $2^{\otimes 7}$, $2^{\otimes 8}$ only the $D=r=\maxMMErank$ case was tested due to time constraints, but in all cases the MME behavior was confirmed.

\Table{5} strongly suggests that for \textit{even} numbers $N$ of qubits,  $\maxMMErank$ saturates the upper limit $\maxMMEranklim$, so
\begin{Equation}                      {81}
\begin{array}{*{20}c}
   {\maxMMErank=\maxMMEranklim = 2^{N-2}\,;} &\; {\left(\! {\begin{array}{*{20}l}
   {\text{for $N\!\geq\! 2$ qubits,}}  \\
   {\text{when $N$ is even}}  \\
\end{array}}\! \right)\!,}  \\
\end{array}
\end{Equation}
while for \textit{odd} $N$, $\maxMMErank<\maxMMEranklim$ and needs to be found by using \Eq{5} and searching ME TGX tuples as usual. However, \Table{5} shows that the odd-$N$ $\maxMMErank$ appears to be gaining on $\maxMMEranklim$ as $N$ increases.
\begin{table}[H]
\caption{\label{tab:5}Examples of $N$-qubit MME states up to maximal MME rank $\maxMMErank$, where $\rho_{\MME}$ is an MME TGX state with eigenstates represented as ME TGX tuples in scalar-index form, as in \Table{2}. Such states are MME for all spectra of all ranks $r\leq\maxMMErank$ and EPU variations. The notation $\maxMMErank=22^{\ddagger}$ for $2^{\otimes 7}$ denotes that this is merely the largest value of $\maxMMErank$ found \textit{so far}, as explained in the main text.}
\begin{ruledtabular}
\begin{tabular}{|c|c|c|c|c|}
{ $n$ } & { ${\kern -4pt}2^{\otimes N}{\kern -4pt}\rule{0pt}{10pt}$ } & { ${\kern -5pt}{\scriptstyle\maxMMEranklim}{\kern -4pt}$ } & { ${\kern -5pt}{\scriptstyle\maxMMErank}{\kern -4pt}$ } & { Levels of Eigenstates of $\rho_{\MME}$ }\\[0.5mm]
\hline
{ $4$\rule{0pt}{10pt} } & { ${\kern -6pt}2^{\otimes 2}{\kern -4pt}$ } & $1$ & $1$ & { $\{1,4\}$ } \\[0.5mm]%
\hline
{ $8$\rule{0pt}{10pt} } & { ${\kern -6pt}2^{\otimes 3}{\kern -4pt}$ } & $2$ & $1$ & { $\{1,8\}$ } \\[0.5mm]%
\hline
{ $16$\rule{0pt}{10pt} } & { ${\kern -6pt}2^{\otimes 4}{\kern -4pt}$ } & $4$ & $4$ & { $\{1,16\}, \{4,13\}, \{6,11\}, \{7,10\}$ } \\[0.5mm]%
\hline
$\begin{array}{*{20}c}
   {32 \rule{0pt}{10pt}}  \\
   {}  \\
\end{array}$ & $\begin{array}{*{20}c}
   {{\kern -6pt}2^{\otimes 5}\rule{0pt}{10pt}{\kern -4pt}}  \\
   {}  \\
\end{array}$ & $\begin{array}{*{20}c}
   {8 \rule{0pt}{10pt}}  \\
   {}  \\
\end{array}$ & $\begin{array}{*{20}c}
   {5 \rule{0pt}{10pt}}  \\
   {}  \\
\end{array}$ & $\begin{array}{*{20}l}
   { \{1,32\},\{4,29\},\{6,27\},\{10,23\}, }  \\
   { \{15,18\} }  \\
\end{array}$ \\[0.5mm]
\hline
$\begin{array}{*{20}c}
   {64 \rule{0pt}{10pt}}  \\
   {}  \\
   {}  \\
   {}  \\
\end{array}$ & $\begin{array}{*{20}c}
   {{\kern -6pt}2^{\otimes 6}\rule{0pt}{10pt}{\kern -4pt}}  \\
   {}  \\
   {}  \\
   {}  \\
\end{array}$ & $\begin{array}{*{20}c}
   {16 \rule{0pt}{10pt}}  \\
   {}  \\
   {}  \\
   {}  \\
\end{array}$ & $\begin{array}{*{20}c}
   {16 \rule{0pt}{10pt}}  \\
   {}  \\
   {}  \\
   {}  \\
\end{array}$ & $\begin{array}{*{20}l}
   { \{1,64\},\{4,61\},\{6,59\},\{7,58\}, }  \\
   { \{10,55\},\{11,54\},\{13,52\},\{16,49\}, }  \\
   { \{18,47\},\{19,46\},\{21,44\},\{24,41\}, }  \\
   { \{25,40\},\{28,37\},\{30,35\},\{31,34\} }  \\
\end{array}$ \\[0.5mm]
\hline
${\kern -3pt}\begin{array}{*{20}c}
   {128 \rule{0pt}{10pt}}  \\
   {}  \\
   {}  \\
   {}  \\
   {}  \\
   {}  \\
\end{array}{\kern -3pt}$ & $\begin{array}{*{20}c}
   {{\kern -6pt}2^{\otimes 7}\rule{0pt}{10pt}{\kern -4pt}}  \\
   {}  \\
   {}  \\
   {}  \\
   {}  \\
   {}  \\
\end{array}$ & $\begin{array}{*{20}c}
   {32 \rule{0pt}{10pt}}  \\
   {}  \\
   {}  \\
   {}  \\
   {}  \\
   {}  \\
\end{array}$ & $\begin{array}{*{20}c}
   {22^{\ddagger} \rule{0pt}{10pt}}  \\
   {}  \\
   {}  \\
   {}  \\
   {}  \\
   {}  \\
\end{array}$ & ${\kern -1.3pt}\begin{array}{*{20}l}
   { \{1,128\},\{4,125\},\{6,123\},\{7,122\}, }  \\
   { \{10,\!119\},\!\{11,\!118\},\!\{13,\!116\},\!\{18,\!111\}, }  \\
   { \{19,\!110\},\!\{21,\!108\},\!\{25,\!104\},\!\{32,\!97\}, }  \\
   { \{34,95\},\{35,94\},\{37,92\},\{41,88\}, }  \\
   { \{48,81\},\{49,80\},\{56,73\},\{60,69\}, }  \\
   { \{62,67\},\{63,66\} }  \\
\end{array}{\kern -1.3pt}$ \\[0.5mm]
\hline
${\kern -3pt}\begin{array}{*{20}c}
   {256 \rule{0pt}{10pt}}  \\
   {}  \\
   {}  \\
   {}  \\
   {}  \\
   {}  \\
   {}  \\
   {}  \\
   {}  \\
   {}  \\
   {}  \\
   {}  \\
   {}  \\
   {}  \\
\end{array}{\kern -3pt}$ & $\begin{array}{*{20}c}
   {{\kern -6pt}2^{\otimes 8}\rule{0pt}{10pt}{\kern -4pt}}  \\
   {}  \\
   {}  \\
   {}  \\
   {}  \\
   {}  \\
   {}  \\
   {}  \\
   {}  \\
   {}  \\
   {}  \\
   {}  \\
   {}  \\
   {}  \\
\end{array}$ & $\begin{array}{*{20}c}
   {64 \rule{0pt}{10pt}}  \\
   {}  \\
   {}  \\
   {}  \\
   {}  \\
   {}  \\
   {}  \\
   {}  \\
   {}  \\
   {}  \\
   {}  \\
   {}  \\
   {}  \\
   {}  \\
\end{array}$ & $\begin{array}{*{20}c}
   {64 \rule{0pt}{10pt}}  \\
   {}  \\
   {}  \\
   {}  \\
   {}  \\
   {}  \\
   {}  \\
   {}  \\
   {}  \\
   {}  \\
   {}  \\
   {}  \\
   {}  \\
   {}  \\
\end{array}$ & $\scalemath{0.87}{\begin{array}{*{20}l}
   { \{1,256\},\{4,253\},\{6,251\},\{7,250\},\rule{0pt}{10pt} }  \\
   { \{10,247\},\{11,246\},\{13,244\},\{16,241\}, }  \\
   { \{18,239\},\{19,238\},\{21,236\},\{24,233\}, }  \\
   { \{25,232\},\{28,229\},\{30,227\},\{31,226\}, }  \\
   { \{34,223\},\{35,222\},\{37,220\},\{40,217\}, }  \\
   { \{41,216\},\{44,213\},\{46,211\},\{47,210\}, }  \\
   { \{49,208\},\{52,205\},\{54,203\},\{55,202\}, }  \\
   { \{58,199\},\{59,198\},\{61,196\},\{64,193\}, }  \\
   { \{66,191\},\{67,190\},\{69,188\},\{72,185\}, }  \\
   { \{73,184\},\{76,181\},\{78,179\},\{79,178\}, }  \\
   { \{81,176\},\{84,173\},\{86,171\},\{87,170\}, }  \\
   { \{90,167\},\{91,166\},\{93,164\},\{96,161\}, }  \\
   { \{97,\!160\},\{100,\!157\},\{102,\!155\},\{103,\!154\}, }  \\
   { \{106,\!151\},\!\{107,\!150\},\!\{109,\!148\},\!\{112,\!145\}, }  \\
   { \{114,\!143\},\!\{115,\!142\},\!\{117,\!140\},\!\{120,\!137\}, }  \\
   { \{121,\!136\},\!\{124,\!133\},\!\{126,\!131\},\!\{127,\!130\} }  \\
\end{array}}$ \\[0.5mm]
\end{tabular}
\end{ruledtabular}
\end{table}
In all systems shown in these tables, we found that all decompositions of candidate MME states yielded an average entanglement of $E=1$ with each system generating plots identical to the MME section of \Fig{1}, giving strong evidence in favor of the proof's claim that these are truly MME states, for which their minimum average entanglement over all decompositions is $E=1$.

The algorithm used to implement \Eq{5} used an exhaustive test that searched all possible sets of unique ME TGX tuples given a particular initial ME TGX tuple, and follows the steps from \Sec{V.A} and \Sec{V.B}. If $\maxMMEranklim$ was reached before the end of the search, the routine was ended early. A faster nonexhaustive routine was used when it achieved $\maxMMErank=\maxMMEranklim$, but if $\maxMMErank<\maxMMEranklim$, results of nonexhaustive tests are not conclusive. This is the case for $2^{\otimes 7}$ in \Table{5}; its reported $\maxMMErank$ is merely a highest MME-compatible rank found \textit{so far} (but it should be noted that the same routine found the correct $\maxMMErank$ for $2^{\otimes 5}$ as verified by the exhaustive algorithm).
\section{\label{sec:VI}Conclusions}
\vspace{-8pt}
This paper has successfully established the existence and characterization of multipartite MME states.  The conditions for MME are encapsulated in \Eq{5} which gives the maximal MME rank $\maxMMErank$ in terms of ME TGX states and the generally loose upper limit $\maxMMEranklim$. Thus, we have generalized the bipartite results of \cite{LZFF} to \textit{all possible} multipartite finite discrete quantum systems. We use $\maxMMErank$ as an \textit{indicator} of a quality we call \textit{MME potential}, which is a system's ability to host MME states. If $\maxMMErank\geq 2$, a system can have MME, but if $\maxMMErank= 1$, then only \textit{pure} ME states can have entanglement $1$.

A major ingredient to proving \Eq{5} is the \textit{multipartite Schmidt decomposition} (MSD), first presented in \cite{HedE}, which \Sec{III} reviews.  Since the MSD also depends on \textit{true-generalized X} (TGX) states \cite{HedX,HedD,MeMG,MeMH,HedE,HedC,HeXU,HCor}, \Sec{III} reviews those as well, and explains their role in the MSD.

\Section{IV} then derives necessary and sufficient conditions for multipartite MME, in analogy to the bipartite case, replacing regular Schmidt decomposition with the MSD (which \textit{is} regular Schmidt decomposition in the bipartite case). Thus \Sec{IV} proves the validity of \Eq{5} and the conditions for the existence of multipartite MME.

Since \Eq{5} is essentially a discrete optimization problem, \Sec{V} shows how to apply \Eq{5} in a practical way, first for a bipartite system for simplicity, and then for a quadripartite system for a full multipartite example. These examples yielded \Fig{1}, \Fig{2}, and \Fig{3}, all of which perform necessary tests to check the validity of \Eq{5}.  For every candidate MME state, we were not able to find any decompositions with a lower average entanglement than $1$, which is the defining feature of MME states.  Thus, while such a test is not exhaustive and therefore not a proof itself, the fact that no such tests found any evidence contradicting the proof in \Sec{IV}, while instead all tests showed only agreement with it, is very strong evidence that the proof in \Sec{IV} is valid.

\Section{V.C} then provides a large number of tabulated results spread out among five tables.  This gives examples of $\maxMMErank$ for many systems. This section also provides ready-to-use examples of sets of ME TGX tuples which are MME compatible, meaning any subset of them can be used with any spectrum to form an MME state up to rank $\maxMMErank$.  This spectrum independence is powerful because it means that entanglement of MME states stays $1$ for all power-sum moments $P_q \equiv \sum\nolimits_{k=1}^{n}\lambda_{k}^{q}=\sum\nolimits_{k=1}^{r}\lambda_{k}^{q}$ for $q\in 1,\ldots,n$ limited only by rank $r\leq\maxMMErank$, such as purity $P\equiv P_{2}$.  Thus, MME states maintain entanglement $1$ down to minimum power-sum values \smash{$P_{q,\min} \equiv r\frac{1}{r^q}|_{r=\maxMMErank}=1/R_{\MME}^{\,q-1}$}.  Interestingly,\hsp{0.7} from\hsp{0.7} \Table{5}\hsp{0.7} and\hsp{0.7} \Eq{81} we see that for even numbers of $N$ qubits, this means that MME states exist down to $P_{q,\min} =\frac{1}{2^{(N-2)(q-1)}}$, which shows that for MME states of these systems,  $\text{lim}_{N\to\infty}( P_{q\geq 2,\min})\to 0$. Therefore, the more qubits there are in a system of even numbers of qubits, the more mixed MME states are allowed to become.  (Although since there can never be an infinite number of qubits, MME of maximally mixed states is never physically attainable, so this does \textit{not} mean that a maximally mixed state can ever have entanglement $1$. Furthermore, such a system would have infinitely many levels $n=\infty$, and therefore would not be a finite system so the methods here would not apply in their given form.)

Since we began with an example of bipartite MME in $2\times 4$ in \Eq{1}, here is an example of multipartite MME for $2\times 2\times 2\times 2$ (from \Table{2}),
\vspace{-3pt}
\begin{Equation}                      {82}
\rho _{\MME}  =\! \frac{1}{2}\scalemath{0.78}{\left( {\begin{array}{*{16}c}
   {\textstyle \lambda _1 } &  \cdot  &  \cdot  &  \cdot  &  \cdot  &  \cdot  &  \cdot  &  \cdot  &  \cdot  &  \cdot  &  \cdot  &  \cdot  &  \cdot  &  \cdot  &  \cdot  & {\textstyle \lambda _1 }  \\
    \cdot  &  \cdot  &  \cdot  &  \cdot  &  \cdot  &  \cdot  &  \cdot  &  \cdot  &  \cdot  &  \cdot  &  \cdot  &  \cdot  &  \cdot  &  \cdot  &  \cdot  &  \cdot   \\
    \cdot  &  \cdot  &  \cdot  &  \cdot  &  \cdot  &  \cdot  &  \cdot  &  \cdot  &  \cdot  &  \cdot  &  \cdot  &  \cdot  &  \cdot  &  \cdot  &  \cdot  &  \cdot   \\
    \cdot  &  \cdot  &  \cdot  & {\textstyle \lambda _2 } &  \cdot  &  \cdot  &  \cdot  &  \cdot  &  \cdot  &  \cdot  &  \cdot  &  \cdot  & {\textstyle \lambda _2 } &  \cdot  &  \cdot  &  \cdot   \\
    \cdot  &  \cdot  &  \cdot  &  \cdot  &  \cdot  &  \cdot  &  \cdot  &  \cdot  &  \cdot  &  \cdot  &  \cdot  &  \cdot  &  \cdot  &  \cdot  &  \cdot  &  \cdot   \\
    \cdot  &  \cdot  &  \cdot  &  \cdot  &  \cdot  & {\textstyle \lambda _3 } &  \cdot  &  \cdot  &  \cdot  &  \cdot  & {\textstyle \lambda _3 } &  \cdot  &  \cdot  &  \cdot  &  \cdot  &  \cdot   \\
    \cdot  &  \cdot  &  \cdot  &  \cdot  &  \cdot  &  \cdot  & {\textstyle \lambda _4 } &  \cdot  &  \cdot  & {\textstyle \lambda _4 } &  \cdot  &  \cdot  &  \cdot  &  \cdot  &  \cdot  &  \cdot   \\
    \cdot  &  \cdot  &  \cdot  &  \cdot  &  \cdot  &  \cdot  &  \cdot  &  \cdot  &  \cdot  &  \cdot  &  \cdot  &  \cdot  &  \cdot  &  \cdot  &  \cdot  &  \cdot   \\
    \cdot  &  \cdot  &  \cdot  &  \cdot  &  \cdot  &  \cdot  &  \cdot  &  \cdot  &  \cdot  &  \cdot  &  \cdot  &  \cdot  &  \cdot  &  \cdot  &  \cdot  &  \cdot   \\
    \cdot  &  \cdot  &  \cdot  &  \cdot  &  \cdot  &  \cdot  & {\textstyle \lambda _4 } &  \cdot  &  \cdot  & {\textstyle \lambda _4 } &  \cdot  &  \cdot  &  \cdot  &  \cdot  &  \cdot  &  \cdot   \\
    \cdot  &  \cdot  &  \cdot  &  \cdot  &  \cdot  & {\textstyle \lambda _3 } &  \cdot  &  \cdot  &  \cdot  &  \cdot  & {\textstyle \lambda _3 } &  \cdot  &  \cdot  &  \cdot  &  \cdot  &  \cdot   \\
    \cdot  &  \cdot  &  \cdot  &  \cdot  &  \cdot  &  \cdot  &  \cdot  &  \cdot  &  \cdot  &  \cdot  &  \cdot  &  \cdot  &  \cdot  &  \cdot  &  \cdot  &  \cdot   \\
    \cdot  &  \cdot  &  \cdot  & {\textstyle \lambda _2 } &  \cdot  &  \cdot  &  \cdot  &  \cdot  &  \cdot  &  \cdot  &  \cdot  &  \cdot  & {\textstyle \lambda _2} &  \cdot  &  \cdot  &  \cdot   \\
    \cdot  &  \cdot  &  \cdot  &  \cdot  &  \cdot  &  \cdot  &  \cdot  &  \cdot  &  \cdot  &  \cdot  &  \cdot  &  \cdot  &  \cdot  &  \cdot  &  \cdot  &  \cdot   \\
    \cdot  &  \cdot  &  \cdot  &  \cdot  &  \cdot  &  \cdot  &  \cdot  &  \cdot  &  \cdot  &  \cdot  &  \cdot  &  \cdot  &  \cdot  &  \cdot  &  \cdot  &  \cdot   \\
   {\textstyle \lambda _1} &  \cdot  &  \cdot  &  \cdot  &  \cdot  &  \cdot  &  \cdot  &  \cdot  &  \cdot  &  \cdot  &  \cdot  &  \cdot  &  \cdot  &  \cdot  &  \cdot  & {\textstyle \lambda _1}  \\
\end{array}} \right)}{\kern -1pt},
\end{Equation}
for $r\!=\!\maxMMErank\!=\!4$ (the example in \Sec{V.B} was a rank-$2$ version of this). From \Table{2}, $2\times 8$ also has $\maxMMErank=4$, but its coincidence basis is different, so its MME-compatible ME TGX tuples are generally different too. Note that MME states generally have TGX form (and EPU variations), and do \textit{not} typically have X form; \Eq{1} and \Eq{82} are just special cases of systems where \textit{some} TGX states happen to have X form. In other systems, such as $3\times6$, pure ME TGX states \textit{never} have X form, so MME states would not be convertible to X form since MME states are always convertible to mixtures of phaseless pure ME TGX states.

Now that we have fully proved and demonstrated multipartite MME, we are  obligated to verify that any entanglement measures $E(\rho)$ we employ give the proper MEMS values of $E=1$ for MME states in systems with MME potential.  Therefore this imposes a crucial constraint on how entanglement measures must perform.

The \textit{applications} of multipartite MME are only as limited as our imaginations.  In \cite{LZFF}, it was shown that bipartite MME can be used for quantum teleportation \cite{BBCJ,BPM1,BPM2,HedA}.  Therefore, it is likely that multipartite MME can also be used for many-body quantum teleportation protocols, as well as remote-state preparation \cite{HKLo,Pati,BDSS,ArPP,ACNK,ZhSo,WaZL,PrYa,WZAC,JZLW,WaYG} which generally requires many-body entanglement.  Furthermore, given the importance of multipartite entanglement in other areas, such as quantum cryptography \cite{BB84,B92,Eker} and quantum computing \cite{Feyn,DiVi,Sho1,Sho2,Grov,Deu1,DeJo,CEMM}, multipartite MME is likely to be a valuable resource, with possibilities beyond pure ME.

Since the novel feature of MME states is that they are \textit{not pure}, yet still have entanglement $1$ for any spectrum of an MME-compatible rank, they may have powerful applications as states that maintain perfect entanglement in the presence of certain types of noise, such as quantum channels that only affect spectrum.  Therefore, all of this suggests many exciting areas for future research into applications of multipartite MME states.
%
\end{document}